\documentclass[aps,twocolumn]{revtex4-2}

\usepackage{amsthm} % for definitions and theorems
\usepackage{amsmath}
\usepackage{mathtools}
\usepackage{bbold}  % for the bold style of identity matrices etc...
\usepackage{physics} % for bra-ket notation
\usepackage{xcolor} % for text coloring

\usepackage{adjustbox} % scaling diagrams
\usepackage{centernot}
\usepackage[T1]{fontenc}
\usepackage[normalem]{ulem} % text strikethrough with no redefinition of \emph{}
\usepackage{setspace}
\usepackage[caption=false]{subfig}
\usepackage{float}
\usepackage{graphicx} % multiple figures
\usepackage{optidef} %optimization layout
\usepackage{wrapfig} %Warpfig

\newcommand{\thermomaj}{\succ_{\rm th}}

\newcommand{\gs}[1]{\textcolor{black}{#1}}

\theoremstyle{definition}

\begin{document}

% ### TITLES ###

%\title{Correlations as a resource in molecular switches}
\title{\gs{Correlated dynamics} as a resource in molecular switches}

\author{Daniel Siciliano$^1$}
\email{daniel.siciliano@uni-ulm.de}
%\affiliation{Institute of Theoretical Physics \& IQST, Ulm University, Albert-Einstein-Allee 11, 89081 Ulm, Germany}

\author{Rudi B. P. Pietsch$^1$}
%\altaffiliation{These authors contributed equally to this work}
\email{rudi.pietsch@uni-ulm.de}
% \affiliation{Institute of Theoretical Physics \& IQST, Ulm University, Albert-Einstein-Allee 11, 89081 Ulm, Germany}

\author{Giovanni Spaventa}
\email{giovanni.spaventa@uni-ulm.de}
% \affiliation{Institute of Theoretical Physics \& IQST, Ulm University, Albert-Einstein-Allee 11, 89081 Ulm, Germany}

\author{Susana F. Huelga}
\email{susana.huelga@uni-ulm.de}
% \affiliation{Institute of Theoretical Physics \& IQST, Ulm University, Albert-Einstein-Allee 11, 89081 Ulm, Germany}

\author{Martin B. Plenio}
\email{martin.plenio@uni-ulm.de}
\thanks{\flushleft $^1$ These authors contributed equally to this work}
\affiliation{Institute of Theoretical Physics \& IQST, Ulm University, Albert-Einstein-Allee 11, 89081 Ulm, Germany}

% \author{Daniel Siciliano}
% \altaffiliation{These authors contributed equally to this work}
% \email{daniel.siciliano@uni-ulm.de}
% \affiliation{Institute of Theoretical Physics \& IQST, Ulm University, Albert-Einstein-Allee 11 89081, Ulm, Germany}
% \author{Rudi B. P. Pietsch}
% \altaffiliation{These authors contributed equally to this work}
% \email{rudi.pietsch@uni-ulm.de}
% \affiliation{Institute of Theoretical Physics \& IQST, Ulm University, Albert-Einstein-Allee 11 89081, Ulm, Germany}
% \author{Giovanni Spaventa}
% \email{giovanni.spaventa@uni-ulm.de}
% \affiliation{Institute of Theoretical Physics \& IQST, Ulm University, Albert-Einstein-Allee 11 89081, Ulm, Germany}
% \author{Susana F. Huelga}
% \email{susana.huelga@uni-ulm.de}
% \affiliation{Institute of Theoretical Physics \& IQST, Ulm University, Albert-Einstein-Allee 11 89081, Ulm, Germany}
% \author{Martin B. Plenio}
% \email{martin.plenio@uni-ulm.de}
% \affiliation{Institute of Theoretical Physics \& IQST, Ulm University, Albert-Einstein-Allee 11 89081, Ulm, Germany}
% \date{\today}

\begin{abstract}  
Photoisomerization, a photochemical process underlying many biological mechanisms, has been modeled recently within the quantum resource theory of thermodynamics. This approach has emerged as a promising tool for studying fundamental limitations to nanoscale processes independently of the microscopic details governing their dynamics. On the other hand, correlations between physical systems have been shown to play a crucial role in quantum thermodynamics by lowering the work cost of certain operations. Here, we explore quantitatively how correlations between multiple photoswitches can enhance the efficiency of photoisomerization beyond that attainable for single molecules. Furthermore, our analysis provides insights into the interplay between quantum and classical correlations in these transformations.
\end{abstract}

\maketitle

{\emph{Introduction --}} Nanoscale systems are in general rather difficult to analyze from a thermodynamical point of view, due to the complicated microscopic details that govern their dynamics, and due to their fundamentally out-of-equilibrium nature. However, a recent promising approach is that offered by Quantum Resource Theories (QRTs) \cite{coecke2016mathematical,chitambar2019quantum}, and in particular the formulations of thermodynamics that have emerged from them \cite{ruch1976principle,ruch1978mixing,janzing2000thermodynamic,horodecki2013fundamental,goold2016role,lostaglio2019introductory,ng2018resource}. These frameworks can be used as a toolbox for the study of fundamental limitations to nanoscale processes that would otherwise be difficult to capture by solely relying on classical, equilibrium and macroscopic
frameworks \cite{halpern2020fundamental,spaventa2022capacity,burkhard2023boosting, tiwary2024quantum, alyuruk2025thermodynamic}. In particular, QRTs allow the analysis of a physical process on purely fundamental grounds, without providing a detailed description of the underlying microscopic theory that governs it, e.g. the specific environmental structure, the coupling strengths, etc \gs{\footnote{\gs{We note that some assumptions on the spectrum of the bath are needed, see \cite{horodecki2013fundamental}. However, as the authors discuss, these are naturally satisfied by e.g. a product \(\tau^{\otimes N}\) of many independent Gibbs states.}}}. A notable example is that of photoisomerization, a photochemical process that is at the basis of human vision \cite{telegina2023isomerization}, plays a key role in the primary steps of photosynthesis in plants, algae and bacteria \cite{croce2018light}, and can also be artificially controlled for technological applications, such as the storage of solar energy, nanorobotics and optical data storage \cite{dattler2019design}. Crucially, the microscopic details of the physics of photoisomerization are rather difficult to capture due to its non-equilibrium nature, its ultra-fast speed, the involvement of vibrational modes, and its very high quantum yield \cite{nogly2018retinal,seidner1994microscopic,seidner1995nonperturbative,hahn2000quantum,hahn2002ultrafast}. For this reason, recent works \cite{halpern2020fundamental,spaventa2022capacity,burkhard2023boosting} have deployed the resource theory of athermality to find fundamental limitations to the efficiency of photoisomerization, in a way that is independent of the microscopic details governing the dynamics. As already proposed and explored in those works and in \cite{DanielThesis, RudiThesis}, the photoisomerization yield of two molecules can improve over that of a single one, which makes it natural to ask how this advantage behaves for larger ensembles of molecules, and in particular how the $N$ molecule solution approaches that found in the thermodynamic limit, via the standard second law of thermodynamics. Indeed, the role of correlations in quantum thermodynamics is nontrivial. On the one hand, the thermal state of a system might be correlated or not depending on the particular choice of Hamiltonian, implying that the resourcefulness of correlated states depends on the details of the system (e.g. thermal states of non-interacting systems do not exhibit correlations, therefore any correlated state will represent a resource). On the other hand, quantifying the impact of correlations on state conversion criteria is not straightforward, due to \gs{their interplay with resources such as energy and purity}. Nevertheless, it has been proven that correlations can allow some thermodynamical protocols to be implemented at a lower work cost \gs{\cite{sapienza2019correlations, Jennings.PhysRevE.81.061130,partovi2008entanglement, de2020temporal}}. \gs{An example of this is imperfect catalysis, where a system can become correlated to a catalyst after undergoing a joint transformation \cite{lostaglio2015stochastic, rubboli2022fundamental}}. \\ 
Here, we want to quantitatively explore the impact of correlations between different molecules on the efficiency of photoisomerization. We do so by explicitly computing the optimal photoisomerization yield in a many-molecule scenario under the assumption that the molecules end up in a correlated or uncorrelated state, and we compare the solutions. We find that \gs{allowing for (classical) correlations to be generated} offers a significant boost in efficiency, and we can compute how this advantage scales with the number of molecules. Furthermore, we quantify the role of quantum correlations \gs{in the initial state} on the photoisomerization yield of two molecules and find a small advantage over classical correlations.\\ 

{\emph{Outline --}} The paper is organized as follows. First, we briefly introduce the resource theory of thermodynamics and the study of photoisomerization within this framework to fix the notation. \gs{Then, we define the \textit{thermodynamic limit}, where the number of photoswitches becomes large}. Once the relevant quantities that we want to study, such as the process yield, are defined, we analyze the role of correlations when two molecules undergo a photoisomerization process together. As we find that the yield increases compared to the single molecule case, we extend our analysis to the $N$-molecule scenario and the \gs{regime} for which the number of molecules becomes large. Finally, we show that the same tools we developed to study the role of classical correlations \gs{in the final state} can also be applied to coherence \gs{in the initial state}, and we compare the effect of the two in the case of \(N=2\) photoswitches.\\

{\emph{Thermodynamics as a quantum resource theory --}} The framework of quantum resource theories (QRTs) allows the rigorous quantification of physical resources (such as quantum coherence, entanglement, non-Markovianity, etc.) and their interconversion. They provide a theoretical framework in which a set of operations (i.e. a subset of all quantum channels) are considered \textit{free}, and any state that cannot be prepared via free operations is then singled out as a \textit{(static) resource}, in the sense of facilitating a task inaccessible to the free set. Non-free states (or operations) can thus only be prepared (or implemented) at a cost, while on the other hand assisting processes that would be otherwise impossible or only attainable with a smaller fidelity. Historically, the first example of a resource theory is the theory of bipartite entanglement \cite{plenio2007introduction,horodecki2009quantum}, where the restriction to local operations and classical communication (LOCC) promotes entanglement to a resource, while separable states remain freely accessible. Analogously, quantum thermodynamics can be formulated as a resource theory in which both free states and operations are \emph{thermal} \cite{ruch1976principle,ruch1978mixing,janzing2000thermodynamic,horodecki2013fundamental,goold2016role,lostaglio2019introductory,ng2018resource}.  Specifically, the resource theory of \emph{athermality} is constructed as follows. Given a system $S$ with Hamiltonian $H_S$, the following three elementary operations are allowed: (i) The system can be brought into contact with a thermal bath $B$ at inverse temperature $\beta$, that is, we can freely deploy Gibbs states $\tau = e^{-\beta H_B}/Z$. (ii) We can perform any global unitary transformation $U$ on $S+B$, as long as it is strictly energy preserving, i.e., $ \big[ U,H_S+H_B \big]=0$ . (iii) We are allowed to trace out subsystems, and in particular the entire bath $B$.
As a result, the action of \emph{thermal operations} (\(\mathsf{TO}\)) on a density operator $\rho_S$ is then defined as
\begin{equation} \rho_S \,\xrightarrow{\,\,\mathsf{TO}\,\,}\, \Tr_{B}\, \big[ U\,\rho_S\otimes\tau\, U^\dagger \big]\,. \end{equation}
Note that thermal operations preserve the Gibbs state of the system $S$, and furthermore they obey \textit{time-translation covariance} (also  called \textit{phase-covariance} or $U(1)$-covariance), i.e. they commute with the free unitary evolution of the system: $\mathcal{T}\circ\mathcal{U}_t=\mathcal{U}_t\circ\mathcal{T}$ for any $\mathcal{T}\in\mathsf{TO}$, \gs{where $\mathcal{U}_t = e^{-iH_St}$}.

The action of thermal operations on block-diagonal states in the energy eigenbasis can be fully characterized. The associated state conversion problem, i.e. deciding whether such a state $\rho$ can be mapped into another state $\sigma$ via thermal operations, can be solved via a particular version of relative majorization called \textit{thermomajorization} \cite{ruch1976principle,ruch1978mixing,horodecki2013fundamental}. This is analogous to how standard majorization characterizes state convertibility under LOCC in the resource theory of entanglement. 
In particular, one associates to a density matrix $\rho$ a curve $f_\rho(x)$ (called \textit{thermomajorization curve}) and, given two density matrices $\rho$ and $\sigma$, it is said that $\rho$ thermomajorizes $\sigma$, i.e. $\rho\thermomaj \sigma$, if
$ f_\rho (x) \geq f_\sigma (x) \,, \forall x$.
% Then, given two states $\rho$ and $\sigma$:
% \begin{equation} \rho\, \xrightarrow{\mathsf{TO}}\, \sigma\,\iff\, \rho \thermomaj \sigma\,.  \end{equation}
Then, if \(\rho \thermomaj \sigma\), for every \(\epsilon>0\) there is a thermal operation \(\mathcal{T}_\epsilon\) that maps \(\rho\) arbitrarily close to \(\sigma\), that is to say \(\Vert \sigma- \mathcal{T}_\epsilon(\rho)\Vert \leq \epsilon\).
An equivalent tool to assess state convertibility is the existence of a Gibbs-stochastic matrix mapping the diagonal of \(\rho\) to the diagonal of \(\sigma\). These are stochastic matrices whose fixed point is the population vector of the Gibbs state. This tool will be useful when considering a large number of molecules and will be expanded in the corresponding section.\\

\begin{figure}
\centering\includegraphics[width=0.48\textwidth]{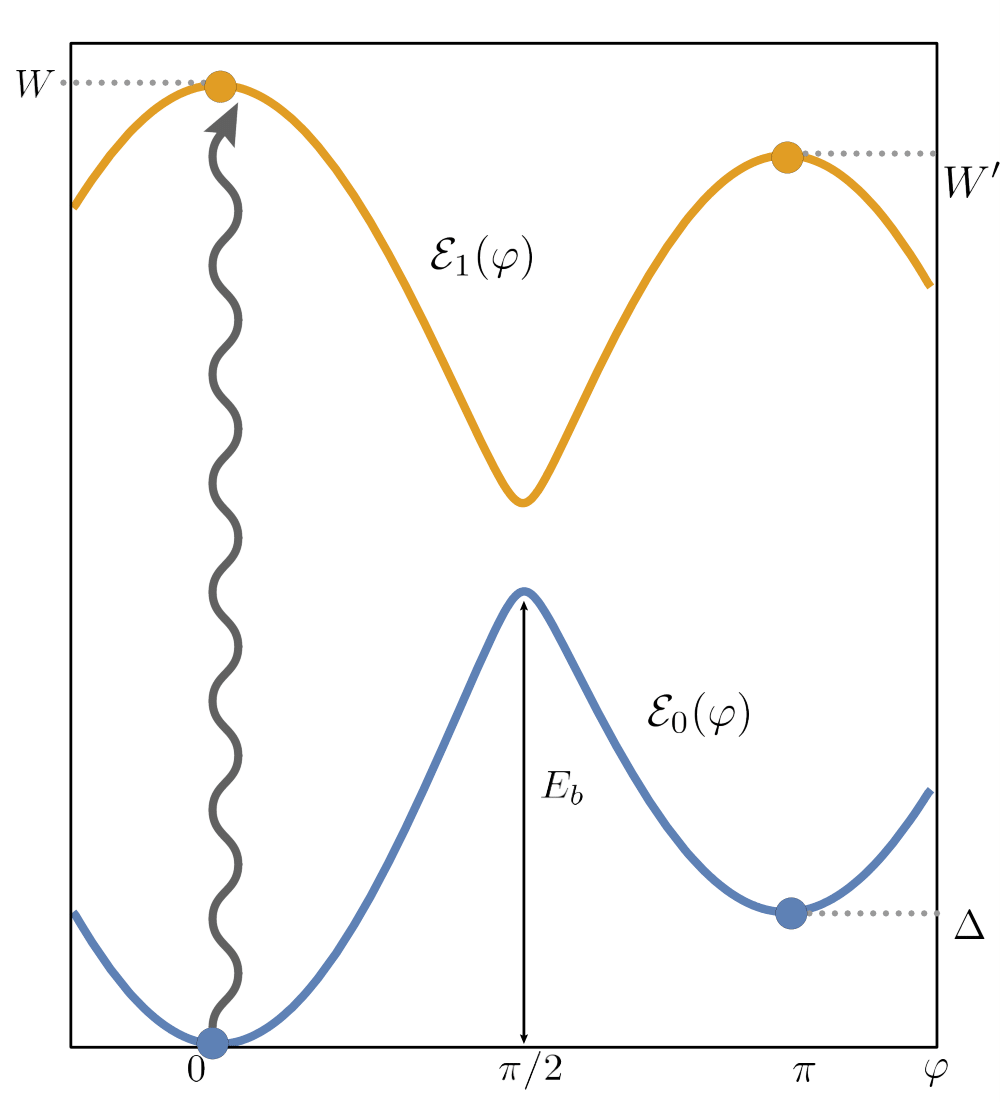}
\caption{Energy landscape for a typical photoisomer. The system starts in the electronic ground state at $\varphi=0$, and it is photoexcited (wavy arrow) by a light source. It can then relax to the cis ground state $\mathcal{E}(\varphi=\pi)$, while in contact with its environment. Our results are independent of the actual intermediate dynamics of the process. The advantage of a resource-theoretic approach, is precisely that of providing general bounds that are valid regardless of the complicated microscopic details governing the evolution of the system. The four dots represent the states that are considered in the four-level model of \cite{halpern2020fundamental}.}
    \label{fig:energy_landscape}
\end{figure}

\emph{Photoisomerization --} Some molecules, when excited by light, undergo a configurational change in their geometry. This photochemical process is known as photoisomerization and allows the reversible switching between two or more stable geometrical configurations of a molecule. The typical scenario is that of two stable configurations (called \textit{cis} and \textit{trans} \textit{isomers}), separated by an energy barrier. The problem of modeling photoisomerization in the resource theory of athermality has been originally addressed in \cite{halpern2020fundamental}, where an angular coordinate $\varphi$ between two heavy chemical groups parametrizes the relative rotation of two molecular groups around a double bond. Fig.\ref{fig:energy_landscape} displays a typical energy landscape for these systems, where the two eigenvalues $\mathcal{E}_{0,1}(\varphi)$, \gs{corresponding to eigenvectors $\ket{\mathcal{E}_{0,1}(\varphi)}$,} can be obtained from a class of Hamiltonians commonly used in the study of photoisomerization (see \cite{seidner1994microscopic,hahn2000quantum}). 
The ground state energies for $\varphi=0$ and $\varphi=\pi$ satisfy $\mathcal{E}_{0}(0)\le\mathcal{E}_{0}(\pi)$
in our analysis, and are separated by an energy barrier.
The probability of switching configuration during relaxation is called \textit{photoisomerization yield}. Following the same arguments proposed in \cite{halpern2020fundamental}, and later expanded in \cite{spaventa2022capacity,burkhard2023boosting,tiwary2024quantum}, we consider photoisomerization as a mapping between states of a three-level system, realized by thermal operations. More precisely, we imagine the electronic state after photoexcitation as being described by a density operator $\rho$, on the Hilbert space spanned by the three states $\ket{0}\gs{= \ket{\mathcal{E}_0(0)}}$, $\ket{\Delta} \gs{= \ket{\mathcal{E}_0(\pi)}}$, \gs{and} $\ket{W} \gs{= \ket{\mathcal{E}_1(0)}}$. The initial state $\rho$ is assumed to only have support on the \textit{trans} configuration, and is then transformed via a thermal operation into a final state $\sigma$:
\begin{equation}
    \rho \xrightarrow[]{\mathsf{TO}} \sigma\,.
\end{equation}

The \textit{photoisomerization yield} $\gamma$ is then defined as the weight of the final state $\sigma$ on the \textit{cis} electronic ground state, that is $\gamma = \bra{\Delta} \sigma \ket{\Delta}$.\\
In this single-molecule scenario, where the three levels are non-degenerate, quantum coherence in the energy eigenbasis cannot affect the photoisomerization yield, due to the time-translation covariance of thermal operations. Therefore, the transformation $\rho\xrightarrow{}\sigma$ is possible via thermal operations if and only if the initial state thermomajorizes the final state. This means that not all values of the yield are allowed by the constraints of thermal operations, and that we could find an upper bound to the yield by solving for the largest value of $\gamma$ such that $\rho \thermomaj \sigma$. However, when multiple molecules are considered, their spectrum will acquire degeneracies. This means that the global thermal operation that evolves them simultaneously will in principle be able to turn coherence between degenerate levels into populations, thus allowing for a boosting of the photoisomerization yield \cite{burkhard2023boosting}. To assess the impact of classical and quantum correlations independently, we study this problem by first excluding the presence of coherence, whose role will be explored in the last section.\\ 
Treating the photoisomerization of multiple molecules requires generalizing the definitions adopted for the single-molecule case. In particular, we will consider an i.i.d. approximation for the initial state, i.e.
\begin{equation}
    \rho_N = \rho^{\otimes N}\,,
\end{equation}
where $\rho$ is the single molecule state
\begin{equation}\label{eq:initial_state}
    \rho = (1-q)\ket{0}\bra{0} + q \ket{W}\bra{W}\,.
\end{equation}
\gs{In a realistic scenario, the initial state might present correlations, for instance, due to spatial proximity in the case of biological photoswitches in the retina, or because of the presence of light-induced collective excitations. However, we choose a product state here to isolate the role of correlations generated by thermal operations and their impact on the efficiency of the process.\\
Furthermore, we assume the molecules to be sufficiently excited such that $q \geq 1/(1+e^W)$ (cf. Appendix \ref{App:thermomajorization_curves}).}

As for the final state $\sigma_N$, we can consider two different scenarios, depending on whether such a composite state of $N$ molecules is correlated or not. More precisely, in the \textit{uncorrelated final state} scenario, we consider $\sigma_N$ as a product state, while we relax this assumption in the \textit{correlated final state} scenario. The photoisomerization yield for $N$ molecules is instead defined as the (ensemble) average single-molecule photoisomerization yield 
\footnote{This definition of yield corresponds to the expected number of switched molecules out of the ensemble, which we find to be a natural choice. Other definitions of (two-molecule) yields have been proposed in \cite{burkhard2023boosting}, for example considering the probability that at least one molecule is found to be switched after the process. However, this choice assigns the same weight to processes switching only one molecule or both, which in principle can have very different thermodynamical costs.}:
\begin{equation}
    \label{observable}
    \mathcal{O}  = \frac{1}{N} \sum_{i=1}^N (|\Delta \rangle \langle \Delta |)_i, 
\end{equation}
\begin{equation}
    \gamma_N = \text{Tr}(\mathcal{O}\sigma_N) \, ,
\end{equation}
where the subscript \(i\) indicates an operator acting on the \(i\)-th molecule.
The equations above make it clear that the yield is invariant under permutations of subsystems. The analysis can then be simplified by restricting the set of final states to being symmetric states, i.e. states invariant under permutations. \gs{In fact, one can show that} any state with a non-uniform distribution over degenerate levels thermomajorizes a symmetric state \gs{with the same yield}.
\gs{$\gamma_N$} can be equivalently computed by summing all the contributions \gs{to the yield} coming from states of the form \(\ketbra{\Delta}{\Delta}^{\otimes k}\otimes \ketbra{0}{0}^{\otimes (N-k)}\),
i.e. states in which $k$ molecules are switched \gs{(contributing to the yield with a weight of $k/N$ each)} and $N-k$ molecules are not. Fixing a value of $k$, there are precisely $\binom{N}{k}$ states of this kind, with associated populations $p_k^j$, $j=1,\dots,\binom{N}{k}$. Following the assumption of symmetric states, one has $p_k^j=p_k$ $\forall j$. By summing all the contributions over $k$ we obtain the photoisomerization yield
\begin{equation}
    \gamma_N = \sum_{k=1}^N \binom{N}{k}\frac{k}{N} p_k = \sum_{k=1}^N \binom{N-1}{k-1} p_k \,.
    \label{eq:yield_composition}
\end{equation}\\

\emph{Thermodynamic limit --} 
Let us consider the uncorrelated scenario
\begin{equation}\label{asymptoticsc}
    \rho^{\otimes N} \xrightarrow[]{\mathsf{TO}} \sigma^{ \otimes N} \, ,
\end{equation}

In the asymptotic limit, i.e. when $N\to\infty$, we can use a known result on optimal asymptotic conversion rates \cite{brandao2013resource} to find the optimal uncorrelated yield. Let us consider the state conversion problem
\begin{equation}\label{asymptoticsc}
    \rho^{\otimes N} \xrightarrow[]{\mathsf{TO}} \sigma^{\otimes rN} \, ,
\end{equation}
corresponding to the conversion rate \(r\). When \(N \to \infty\), the optimal conversion rate \(r^*\) is given by
\begin{equation}
    r^*= \frac{D(\rho \mid \mid \tau)}{D(\sigma \mid \mid \tau)} = \frac{F(\rho)-F(\tau)}{F(\sigma)-F(\tau)} \, ,
\end{equation}
where \(D(\rho \mid \mid \tau)\) is the quantum relative entropy of \(\rho\) with respect to the thermal state \(\tau\) and $F(\rho)=\text{Tr}(\rho H) - \frac{1}{\beta} S(\rho)$ is the non-equilibrium free energy.
According to \gs{the permutational invariance of the final state}, the optimal uncorrelated yield is the maximum value of \(\gamma =\text{Tr}(\ketbra{\Delta}{\Delta} \sigma)\) such that \(r^*=1\). Therefore, in this limit, \(F(\rho)=F(\sigma)\) defines the optimal asymptotic uncorrelated yield $\gamma_{TD}$, which is the value of \(\gamma\) that saturates the second law \(F(\rho) \ge F(\sigma)\). This equality can then be rewritten as
\begin{equation}
\begin{split}
    F(\rho) &= q W + q\ln q + (1-q)\ln (1-q) \\
    &= \gamma \Delta + \gamma\ln \gamma + (1-\gamma)\ln (1-\gamma) = F(\sigma) \, .
\end{split}
\end{equation}
It is easy to see that $\gamma_{TD} \equiv 1$ is a valid solution as long as $\Delta \leq \Delta^* \gs{\equiv} F(\rho) $, i.e. as long as the \textit{cis-trans} gap is smaller than the free energy of the initial state. For larger values of $\Delta$, the function $\gamma_{TD}(\Delta)$ starts decaying, although a closed-form expression cannot be obtained.\\

% END

\begin{figure}
\centering
\includegraphics[width=0.5\textwidth]{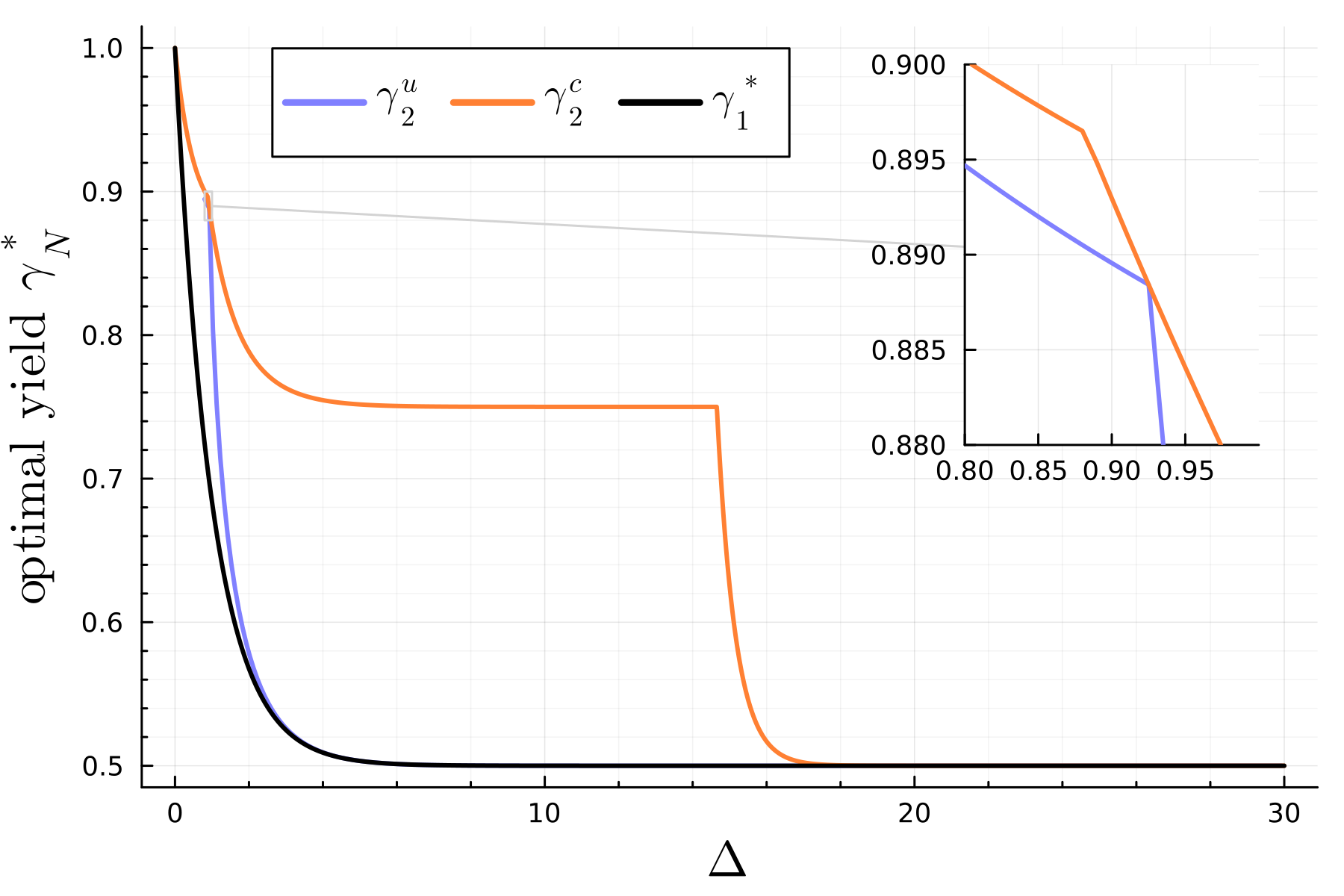}
\caption{Photoisomerization yield for two molecules with uncorrelated and correlated final states. We see that in both cases there is an advantage over the single-molecule yield, which is depicted in black. However, for a final state that is uncorrelated (blue) the yield quickly decreases to the single molecule solution, whereas for a correlated final state (orange), there is a significant improvement of the optimal yield up to $\Delta \approx W/2$. The insert shows a close-up of the small $\Delta$ region where the curves are particularly close and touch at a point. Here, $q=0.5$ and $W=30$.}
\label{fig:N2}
%\end{figure}
%\begin{figure}
%\centering
\includegraphics[width=0.5\textwidth]{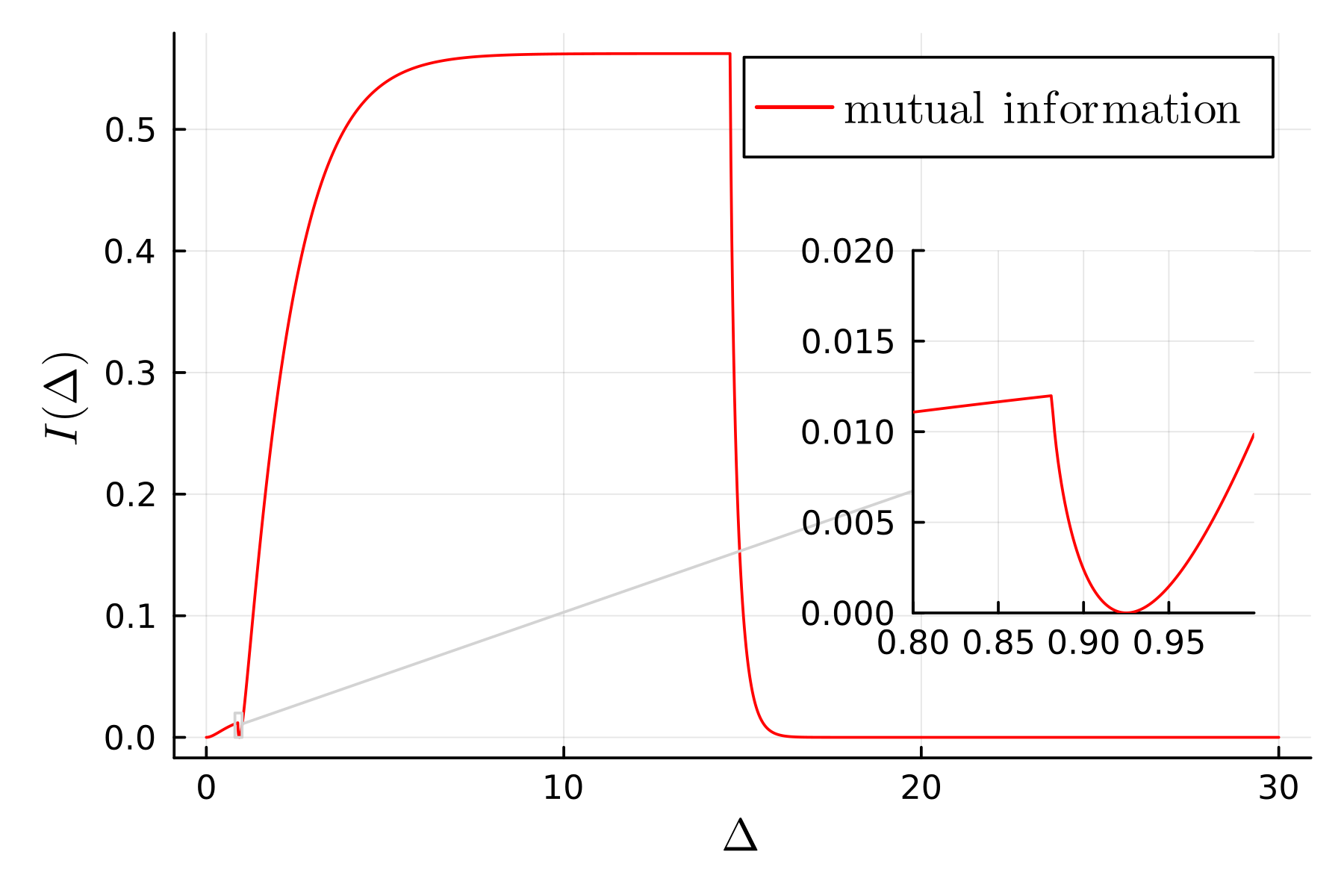}
\caption{Mutual information of the \(N=2\) correlated optimal state. When the correlation advantage becomes larger, the mutual information measures stronger correlations between the two molecules. On the other hand, when the optimal uncorrelated state performs as well as the optimal correlated one, the mutual information approaches zero, indicating that correlations are not needed to reach the optimal yield.}
\label{fig:N2_minfo}
\end{figure}

\emph{Yield optimization on two molecules --}
\label{sec:two_molecules}
As a first step, let us focus on the isomerization of $N=2$ molecules.
In this section, we will compute the optimal yield in the two different scenarios described earlier, i.e., in the case of an uncorrelated and correlated final state.
In the first case, the final state will be a product state of the form
\begin{equation}
    \sigma_2^{u} = \bigl[ \gamma|\Delta\rangle\langle \Delta| + (1-\gamma)|0\rangle\langle 0| \bigr] ^{\otimes 2}\,,
\end{equation}
while in the latter, it will be a general diagonal symmetric state of the form
\begin{equation}
    \begin{split}
    \sigma_2^c & = p_0 |00\rangle \langle00| + p_1 |\Delta 0\rangle\langle\Delta 0| \\
    & + p_1 |0\Delta\rangle \langle 0\Delta | + p_2 |\Delta\Delta\rangle \langle \Delta\Delta|.
    \end{split}
\end{equation}
%\sout{We note that in both cases, regardless of the nature of the final state, correlations are allowed to build up during the \sout{dynamics} switching process. Indeed, if we restricted our model to have no correlations at all times, considering more than one molecule would not affect the optimal yield, as each molecule would act independently and its efficiency would be upper-bounded by the single-molecule yield.}
In the uncorrelated scenario, we note that we still allow a single bath to be coupled with the whole system. There is a third trivial situation in which each molecule undergoes an independent thermal operation, each with its own bath (i.e. the thermal operation has a tensor product structure). The optimal yield will then be upper bounded by the single-molecule optimal yield found in \cite{halpern2020fundamental} and shown in Fig. \ref{fig:N2}.\\ 
To find the optimal yield, we impose the thermomajorization condition between initial and final states using thermomajorization curves. In particular, the curve associated with the final state depends on the populations contributing to the yield. As such, there will be a maximum value of $\gamma$ such that $\rho_2 \succ_{th} \sigma_2^{u,c}$ (see Appendix \ref{App:thermomajorization_curves} for a detailed explanation). In the uncorrelated scenario, the construction of the optimal final curve is simplified by the fact that the populations of the final state depend directly on \(\gamma\). We find the optimal yield $\gamma_2^u$ to be a continuous, piece-wise differentiable function with a single non-differentiable point at the energy value $\Delta^u_{2}$:
\begin{equation}
    \gamma_2^u =
    \begin{cases}
        \sqrt{1-g(q)\left( h(W) - e^{-2 \Delta} \right)}, & \Delta \leq \Delta^u_{2} \\
        1 - \sqrt{g(q)\left( h(W) - e^{-2\Delta} - 2e^{-\Delta} \right)}, & \Delta^u_{2} < \Delta,
    \end{cases}
\end{equation}
where, for the sake of concise notation, we have introduced two auxiliary functions  $g(x) = (1-x)^2$ and $h(x) = (1+e^{-x})^2$.
In the correlated final state scenario, the yield (as defined in Eq. \eqref{eq:yield_composition}) reads
\begin{equation}
    \gamma_2^c = p_2 + p_1\,.
\end{equation} 
As one can see, to find the optimal final curve we now have to optimize over two different parameters. This complication will be more evident when the number of molecules becomes large. Again, the optimal yield is found to be a piecewise differentiable function \gs{(with case distinctions at the $\Delta$-values $\Delta_1^c$ and $\Delta_2^c$)}, of the form 
\begin{equation}\label{eq:correlated_yield_two_molecules}
    \gamma_2^c = \begin{cases}
    1-\frac{1}{2}g(q)\cdot \Big{[} h(W) - e^{-2\Delta} \Big{]}, & \Delta \leq \Delta^c_{1}\\
    1-g(q) \Big{[} \frac{h(W)}{2} - e^{-\Delta}\sqrt{h(\Delta)} \Big{]}, & \Delta^c_{1} < \Delta \leq \Delta^c_{2}\\
    \frac{1}{2} \Bigl\{ 1-g(q)\Big{[} h(W) - h(\Delta) + 1 \Big{]} & \Delta^c_{2} < \Delta, \\ + q^2 + q\sqrt{g(q)}\left( e^{W-2\Delta}-e^{-W} \right) \Bigr\}.
    \end{cases}
\end{equation}
The two solutions and the single molecule optimal yield are shown in Fig. \ref{fig:N2}\gs{, where the plateau of the red curve belongs to the second case of Eq. \eqref{eq:correlated_yield_two_molecules}.} By visual inspection, it is immediately clear that, depending on the value of $\Delta$, the advantage offered by correlations changes dramatically. In particular, there is a regime of $\Delta$ (corresponding to $\Delta \lesssim W/2$) for which this advantage is large. Interestingly, this regime often includes typical photoisomers in nature \gs{(see \cite{wang2021storing, chuang2022steady} and references therein)}. Furthermore, we can see that requiring all correlations to vanish in the final state only gives a slight advantage over the single molecule scenario. We can also measure the correlations in the optimal correlated state by computing the mutual information of the system, considering the reduced density matrices of the two molecules. In Fig. \ref{fig:N2_minfo}, the mutual information is shown to be approaching zero every time the correlation advantage becomes small, such as at large \(\Delta\) and at the point where \(\gamma^u_2=\gamma^c_2\) (\(\Delta \approx 0.925\)). Having this analysis at hand, the question arises of whether this trend continues upon further increasing the number of molecules. \\

\noindent \emph{Yield optimization on a generic number of molecules --}
\label{sec:n_molecules}
When considering $N$ molecules, the uncorrelated final state is given by
\begin{equation}
    \sigma_N^u = \bigl[ \gamma|\Delta\rangle\langle \Delta| + (1-\gamma)|0\rangle\langle 0| \bigr] ^{\otimes N},
\end{equation}
whereas the correlated final state $\sigma^c_N$ has the freedom of being a generic diagonal symmetric state with its population distributed on the same levels as $\sigma_N^u$.
%\begin{equation}
%\begin{split}
%    \rho_f = \sum_{i_1 = 0}^{\Delta} \dddot{} \sum_{i_N = 0}^{\Delta} q_{i_1 \dddot{} \ i_N} & \left( |i_1 \rangle \otimes \dddot{} \ \otimes |i_N \rangle \right) \\ & \cdot \left( \langle i_1 | \otimes \dddot{} \  \otimes \langle i_N | \right) ,
%\end{split}
%\end{equation}
%where we sum over the two possible energy states, i.e. we have $i_k \in \{0, \Delta \} \  \forall k \in [1,...,N]$.
\gs{The optimal yield can be found by constructing and optimizing thermomajorization curves. This results in piecewise functions since the $\beta$-ordering depends on $\Delta$ (see Appendix \ref{App:thermomajorization_curves}).} In the case of two molecules, the number of cases defining the piece-wise function is two for an uncorrelated final state and three for a correlated final state. When we have $N$ molecules, the number of non-differentiable points increases linearly when the final state is uncorrelated and quadratically if we allow for a correlated final state (cf. Appendix \ref{App:thermomajorization_curves}).
%An example of an optimal yield for $N=5$ molecules can be seen in figure XXXXX.
Furthermore, in the correlated case, we optimize over a set of $N$ parameters (cf. Eq. \eqref{eq:yield_composition}) instead of one for the uncorrelated case. Finding upper bounds on the yield for a large number of molecules becomes rapidly impractical as the number of molecules increases, since it requires the construction of numerous thermomajorization curves. Thus, we will now turn to numerical procedures to arrive at meaningful results for the case of large $N$, as well as how the thermodynamic limit to the yield is approached in our model.\\

% \emph{Thermodynamic limit --} Here, the thermodynamic limit is discussed. Uncorrelated thermodynamic yield?

\begin{figure}
\includegraphics[width=\linewidth]{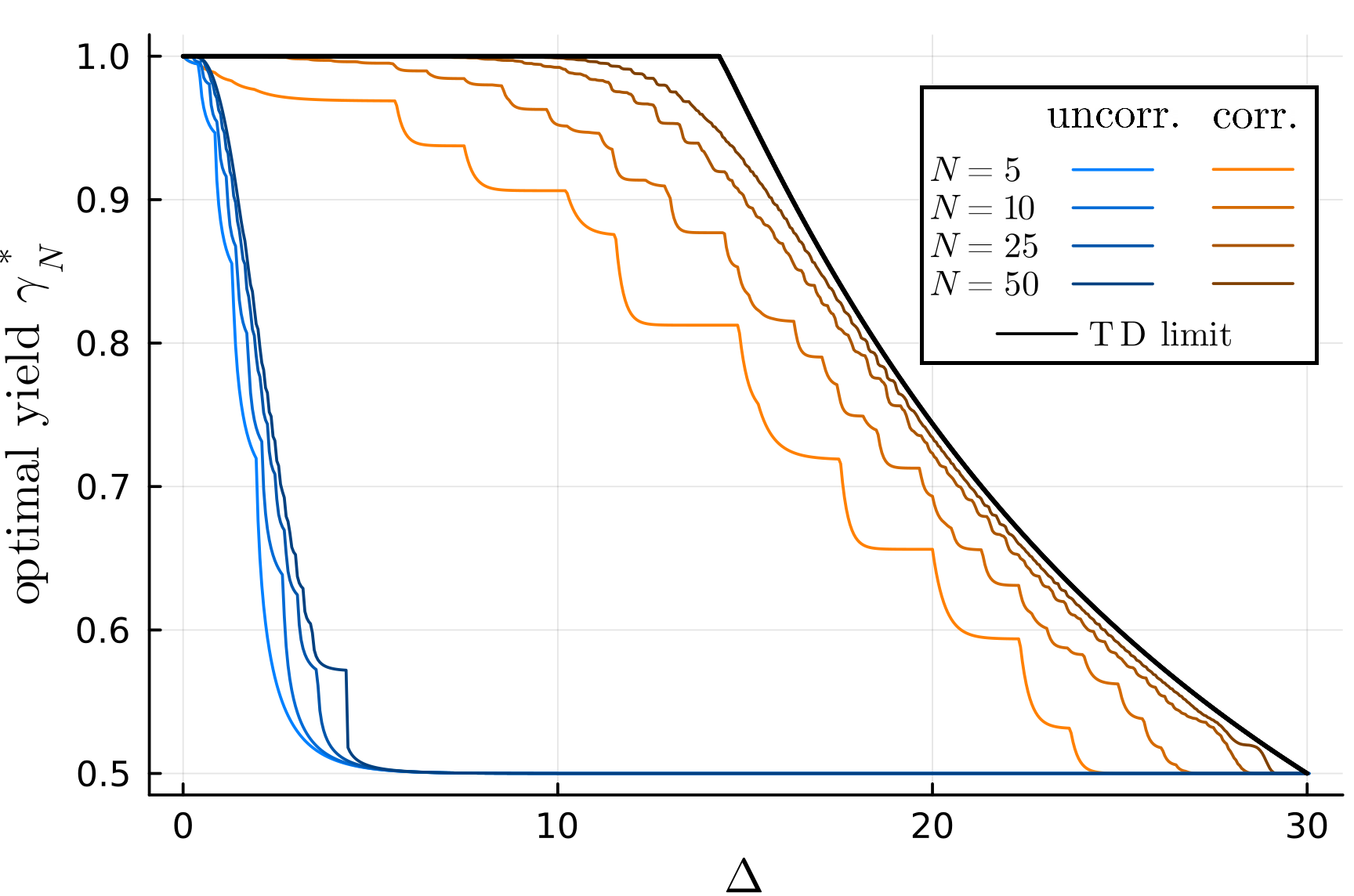}
\caption{Optimal photoisomerization yield for different system sizes. The curves in blue show the maximum yield with an uncorrelated final state hypothesis, while for the orange ones we allow correlations in the final state. The large difference between the two, comparing systems with the same size, shows the positive contribution of correlations on photoisomerization efficiency. The yield is plotted as a function of the energy \(\Delta\) fixing \(q = 0.5\) and \(W = 30\). Appendix \ref{App:GSMprogram} shows a similar plot as a function of \(q\).}
\label{fig:GammaDelta}
\end{figure}

\emph{Numerical approach for multiple photoswitches --} As the number of molecules \(N\) grows large, constructing the thermomajorisation curves to obtain the optimal efficiency for a correlated final state analytically becomes infeasible in practice. Obtaining \(\gamma_N^c\) for larger \(N\) requires a numerical approach, for which we will employ the alternative resource theoretical tool to assess state convertibility: the Gibbs-stochastic matrices ($GS_N$).\\
A matrix \(G \in GS_N\) is a stochastic matrix whose stationary probability vector is the population vector of the Gibbs state:
\begin{equation}
    G p_\tau = p_\tau \, .
\end{equation}
These matrices describe how population vectors are transformed under thermal operations. Our initial and final states are diagonal, so thermomajorization is guaranteed if there is a \(G \in GS_N\) such that:
\begin{equation}
    G p_\rho = p_\sigma \, .
\end{equation}
Our objective is to optimize the efficiency over the $GS_N$ set, given the N-molecule initial state population vector \(p_{\rho_N}\):\\
\begin{equation}\label{eq:optimization}
\max_{G \in GS_N} \Big \{ \gamma(\sigma_N) \mid p_{\sigma_N} = G p_{\rho_N} \Big \}  \, .\\
\end{equation}
The matrix $G$ has dimension \(3^N \times 3^N\), a discouraging scaling with a numerical approach in mind, but we can show that the optimization problem can be reformulated to depend only on a subset of \(N(N+1)\) variables.
\gs{This simplification is made possible by the invariance of the function to be optimized, the yield, under permutations of identical subsystems, and by the simple structure of the initial state, which being a product state of \(N\) identical copies will be characterized by simply \((N+1)\) sets of degenerate, equally populated levels.} A complete treatment of the optimization problem and its variable number reduction, including the role of the constraints, can be found in Appendix \ref{App:GSMprogram}. This pretreatment of the Gibbs-stochastic matrix can be readily adapted to different target functions with the same \gs{invariance} property, possibly outside the context of photoisomerization. The same method is used with \gs{minor} changes in the last section, where coherence enters the picture \gs{in the initial state}, and in Appendix \ref{sec:refining}, where the molecule is modeled as a five-level system.\\
After this simplification, the numerical optimization can be performed efficiently. In Fig. \ref{fig:GammaDelta} we compare the optimal efficiencies for settings up to \(N=50\) molecules. By comparing the optimal correlated and uncorrelated efficiency at fixed \(N\), we appreciate the large advantage provided by allowing for correlations in the final state. Secondly, Fig. \ref{fig:GammaDelta} shows how correlations in the final state allow the maximum efficiency to approach the thermodynamic limit quicker as the number of molecules increases. \\
\\

\begin{figure}
\centering\includegraphics[width=\linewidth]{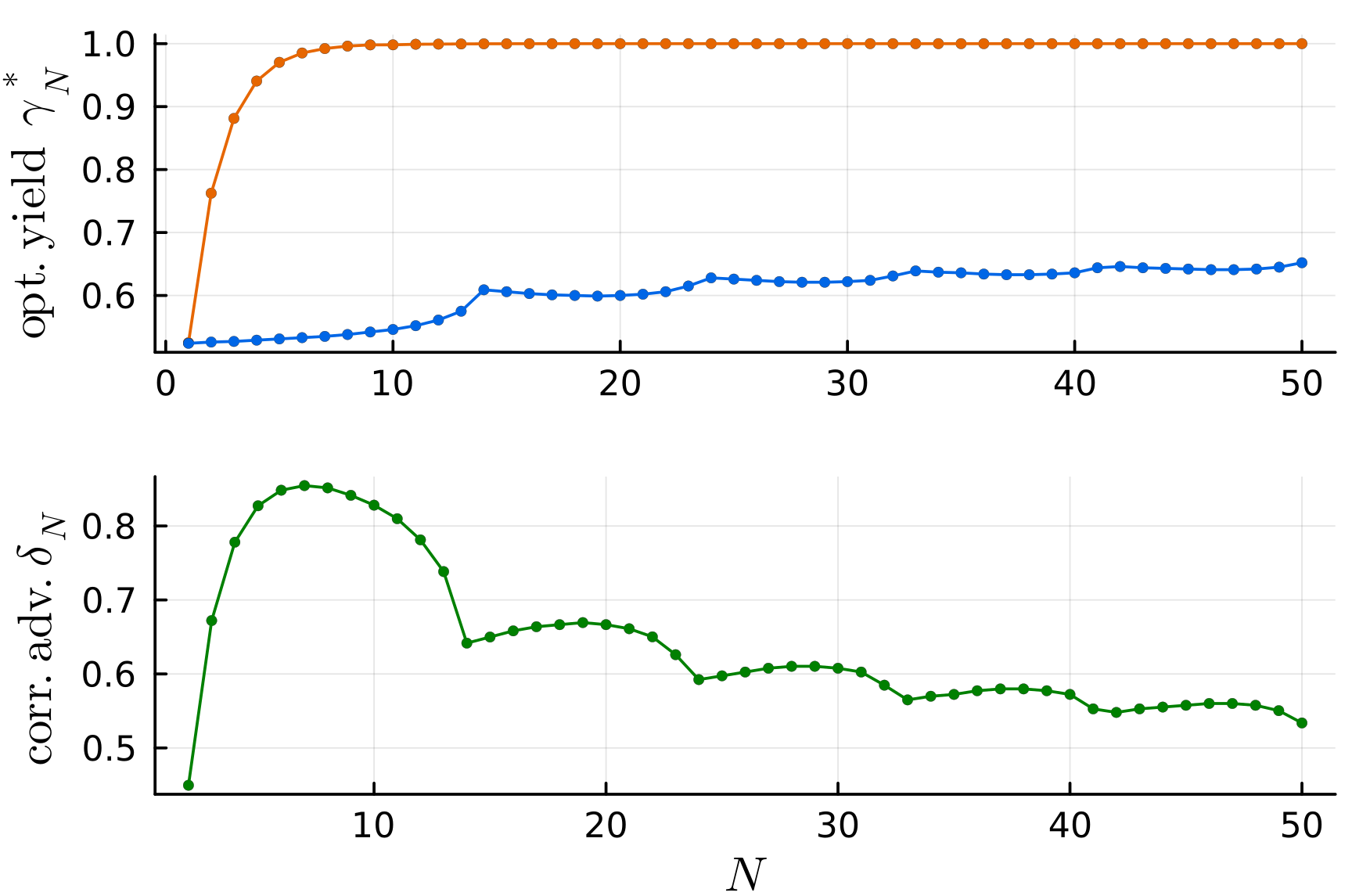}
\caption{The top figure shows the optimal yield as a function of the number of molecules \(N\), for uncorrelated (blue) and correlated (orange) final states. The bottom figure shows the relative difference \(\delta_N = (\gamma_N^c -\gamma_N^u)/\gamma_N^u\). One can notice the non-monotonic behavior of the advantage \(\delta_N\) provided by correlations. The system considered has energies \(\Delta = 3\), \(W = 30\) and initial state population \(q = 0.5\).}
\label{fig:advantage}
\end{figure}

\emph{Correlation advantage and number of photoswitches --} In \cite{sapienza2019correlations}, the advantage provided by correlations in lowering the work of formation of a state is shown to be non-monotonic in the number of copies \(N\), we show here that the same applies to the relative yield advantage
\begin{equation}
    \delta_N = (\gamma_N^c -\gamma_N^u)/\gamma_N^u \, .
\end{equation}
We can now choose a setting, e.g. a system with energies \(\Delta = 3\), \(W = 30\) and initial state population \(q = 0.5\), and then explore the dependency of \(\delta_{N}\) on \(N\). With this choice of parameters, we have that \(\Delta < \Delta^*\) and thus \(\gamma_{TD} = 1\). In Fig. \ref{fig:advantage}, we plot \(\gamma^u_N\), \(\gamma^c_N\) and the relative advantage for this system.
Here, we can appreciate the fast convergence to the thermodynamic limit of the correlated scenario, as opposed to the slow and non-monotonic increase of the uncorrelated optimal yield. We can also observe the relative advantage \(\delta_N\) being non-monotonic and presenting many local maxima. The non-monotonicity of \(\Delta\) closely resembles the non-monotonic behavior of the $C$-work of formation found in \cite{sapienza2019correlations} (Fig. 5, supplemental material).\\
It is clear from Fig. \ref{fig:advantage} that the non-monotonicity of \(\delta_N\) results from the non-monotonicity of \(\gamma_N^u\). This behavior in $N$ of the uncorrelated optimal yield might be surprising, as one might expect that increasing the number of molecules should not be detrimental to the process. Surely, when using $N+1$ molecules, one possible protocol is to isolate one of them and perform a thermal operation that is a tensor product
\begin{align}
    \label{eq:protocol}
    \mathcal{T}=\mathcal{T}_{1,\dots,N}\otimes \mathcal{T}_{N+1}\,,
\end{align}
i.e. a global thermal operation on the first $N$ molecules, and a single copy thermal operation on the last. \gs{However, this intuitive reasoning results in a negative lower bound for the yield difference} of the \(N+1\) and \(N\) molecule systems. Let us define:
\begin{align}
    \frac{\delta\gamma^*_N}{\gamma^*_N} = \frac{\gamma^*_{N+1}-\gamma^*_N}{\gamma^*_N}\,,
\end{align}
which quantifies the relative advantage on the optimal yield $\gamma^*_N$ of $N$ molecules due to the presence of an additional molecule. By using the protocol laid out in Eq. \eqref{eq:protocol}, the optimal yield $\gamma_{N+1}^*$ is guaranteed to be upper-bounded by
\begin{align}
    \gamma^*_{N+1}\geq \frac{N}{N+1}\gamma^*_N + \frac{1}{N+1}\gamma^*_1\,, 
\end{align}
which implies
\begin{align}
    \frac{\delta\gamma^*_N}{\gamma^*_N} \geq -\frac{1}{N+1}\frac{\gamma_N^* - \gamma_1^*}{\gamma^*_N}\,.
    \label{eq:rel_adv}
\end{align}
Now, the relative advantage of $N$ copies over a single copy can be trivially bounded by a protocol implementing the thermal operation
\begin{align}
    \mathcal{T}=\bigotimes_{k=1}^N \mathcal{T}_k\,,
\end{align}
which acts independently on each copy, guaranteeing
\begin{align}
    \gamma^*_N\geq \gamma_1^*\,.
\end{align}
This means that the quantity on the right-hand side of Eq.\eqref{eq:rel_adv} is always negative, and \gs{the} simple protocol \gs{laid out in Eq. \eqref{eq:protocol}} does not guarantee an advantage in using an additional copy. However, this lower bound vanishes for \(N \rightarrow \infty\), imposing monotonicity in the limit where \(N\) grows large.\\

\emph{Large N \gs{regime} --} When the number of molecules becomes large, typicality arguments can be used to find approximations, particularly on the initial product state. This can, in turn, be used to capture the large \(N\) behavior \gs{of} the photoisomerization yield. The initial product state \(\rho^{\otimes N} = \left [  (1-q) \ket{0} \bra{0} + q \ket{W} \bra{W} \right ]^{\otimes N}\) can in particular be approximated by its most populated energy levels. For this state, the total population on a set of degenerate levels with energy \(k W\) is 
\begin{equation}\label{eq:popdistr}
p_k = \binom{N}{k} q^k (1-q)^{N-k}, \quad k = 0...N \, .
\end{equation}
In the limit where $N \rightarrow \infty $, the distribution peaks at the levels closest to \(k \approx qN \) at energy \(E \approx qNW\). We can obtain a reduced description of the initial state by discarding populations away from this energy subspace and then renormalizing the state, thus reducing the number of relevant parameters in the thermomajorization problem. This approach can simplify both the construction of the thermomajorization curves and the optimization over the  $GS_N$ set. The asymptotic limit can be obtained by keeping only the most typical sequences, that is, to keep only the levels \(k = qN\) of energy \(E = qNW\) (with the constraint for \(k\) to be an integer). We can thus use an approximated initial state \(\tilde{\rho}_N\) with zero population everywhere except on those levels, each level having population \(\binom{N}{qN} ^{-1}\).\gs{One can easily show that the relative error on the free energy $\delta F /F :=\big(F(\rho^{\otimes N})-F(\tilde{\rho}_N)\big)/F(\rho^{\otimes N})$ goes to zero since it scales as \( \mathcal{O}(\log{N}/N) \). This can be seen by performing a Stirling approximation on the expression for $F(\tilde{\rho}_N)$.}\\
The linear optimization problem can thus be rewritten to account for this simplification, obtaining a problem that scales only linearly with the number of molecules (see Appendix \ref{App:GSMasymptotic} for details). Moreover, it is possible to show analytically that for \(\Delta \leq \Delta^{*}\) it holds that

\begin{equation}
\gamma_\infty^c(\Delta \leq \Delta^{*}) = \gamma_{TD}(\Delta \leq \Delta^{*}) = 1.
\end{equation}

\begin{figure}
\centering
  \includegraphics[width=\linewidth]{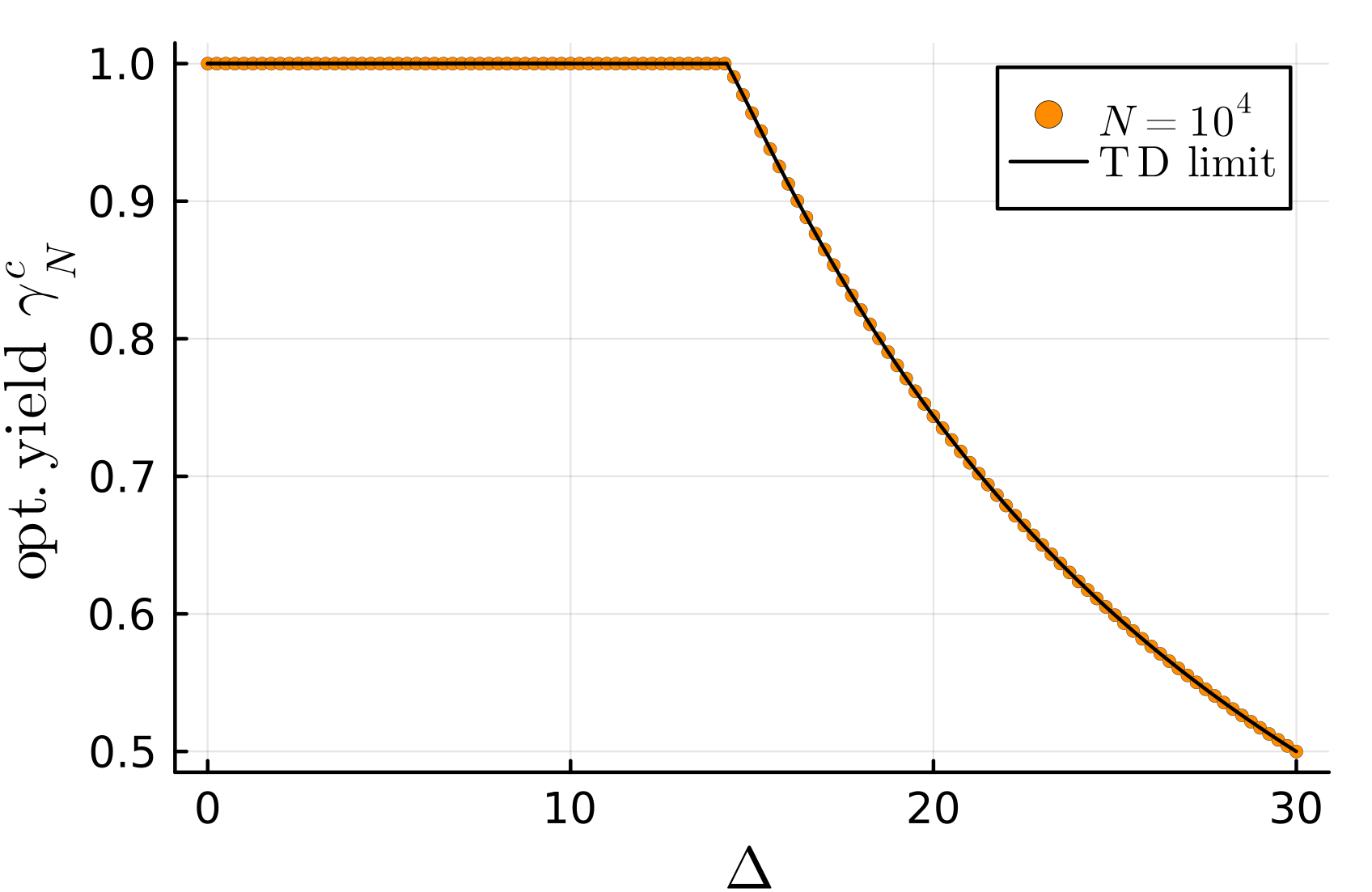}
  \caption{Correlated optimal yield for \(N=10^4\) molecules, with the typical sequences approximation performed on the initial state. The excellent match shown between \(\gamma^c_{N}\) and \(\gamma_{TD}\) when \(N\) is large \gs{suggests} that \(\gamma_{N \rightarrow \infty}^c = \gamma_{TD}\).}
  \label{fig:10000}
\end{figure}

We expect the same to apply for \(\Delta > \Delta^{*}\), and by numerically solving this optimization problem we obtain an excellent numerical match, shown in Fig. \ref{fig:10000}.\\

\begin{figure}
\centering\includegraphics[width=\linewidth]{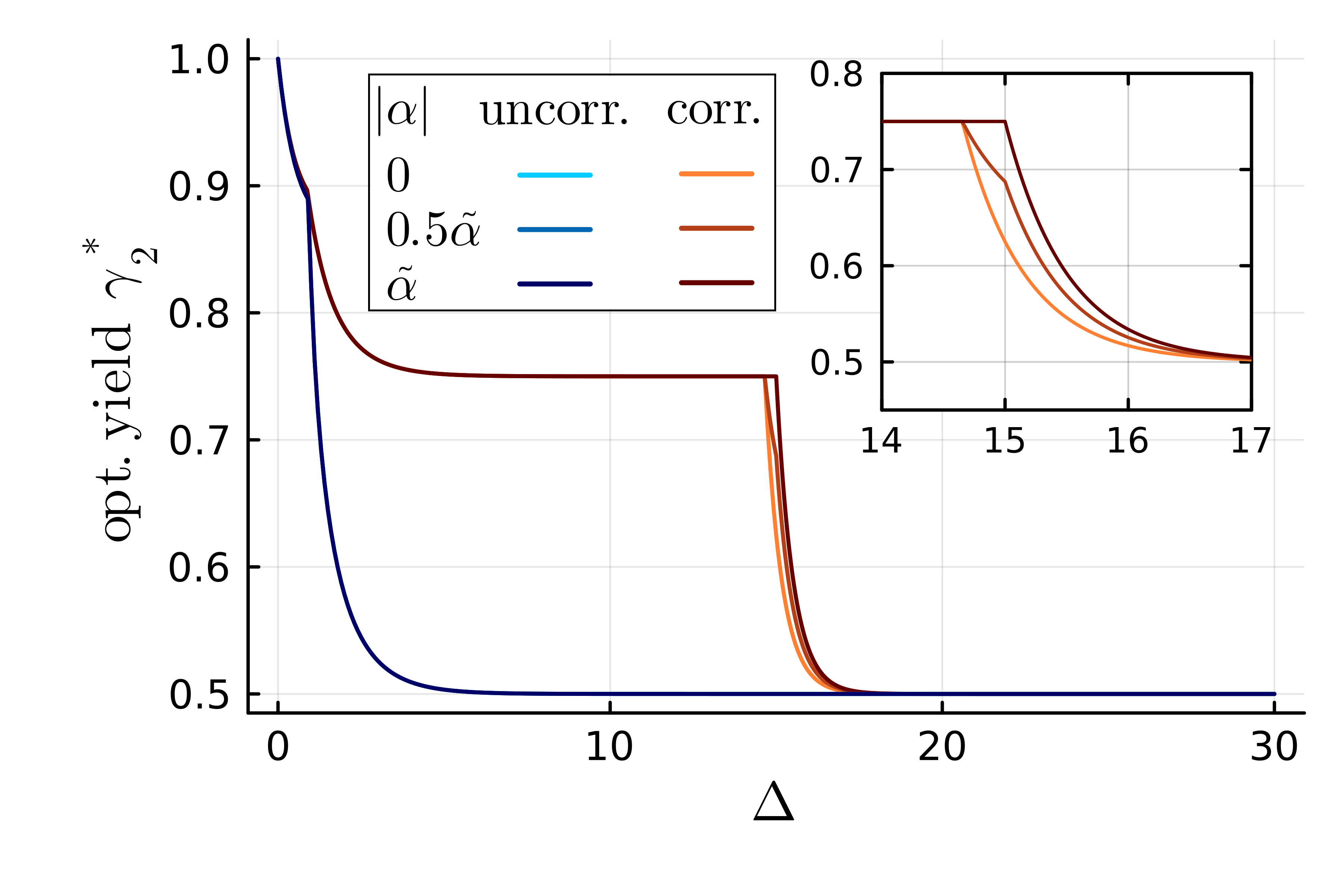}
\caption{Optimal yield for a 2 molecule process in which some coherence is present, \(|\alpha| = \tilde\alpha\) being the maximally coherent scenario. If correlations are not allowed in the final state, there is no noticeable advantage in the optimal yield and the blue curves are superposed. When this restriction is lifted, we see that in a small \(\Delta\) interval coherence provides a noticeable advantage, while for most values of \(\Delta\) the advantage is negligible.}
\label{fig:alpha_advantage}
\end{figure}

\emph{The role of coherence -- } Thermal operations decouple the evolution of different modes of coherence \cite{lostaglio2019introductory}, meaning that if the energy spectrum is non-degenerate populations and coherences evolve separately. Suppose the Hamiltonian is degenerate, like in a multiple-copy system. In that case, an energy-preserving unitary can mix coherence and populations in that subspace. This can reduce the entropy of the initial population vector, increasing its thermodynamical value and therefore allowing a higher optimal yield. \gs{As stated earlier after Eq. \eqref{eq:initial_state}, considering such initial, quantum correlations is of physical interest too, as photoisomers in nature can correlate, e.g. due to spatial proximity.}   Thus, \gs{we will now} explore whether quantum correlations in the initial state can further boost the maximum yield. \\
We will focus on a\gs{n} \(N=2\) molecule system. To isolate the role of coherence in the efficiency boost, we will use an initial state which is a product state, as done before, to which we add some coherence in the relevant subspace, meaning:
\begin{equation}
    \rho_2(\alpha) = \rho^{\otimes 2} + \alpha \ket{W0}\bra{0W} + \alpha^*\ket{0W}\bra{W0} \, .
    \label{Eq:alpha}
\end{equation}
In the subspace with energy \(E=W\), there are two equally populated levels with population \(q(1-q)\). The coherence in this subspace can be expressed via the complex parameter \(\alpha\) with \( |\alpha| \in [0,\Tilde{\alpha}]\) \gs{and} \(\Tilde{\alpha} = q(1-q)\) to ensure positive semidefiniteness.
An energy-preserving unitary can then be applied as a free operation to diagonalize this subspace, mixing populations and coherences so that the new degenerate levels have populations \(q(1-q) \pm |\alpha|\). If \(|\alpha| =\Tilde{\alpha}\), the new state will have this subspace's population localized in a single energy level.
As done in the previous sections, we can then optimize the yield by either allowing or excluding correlations in the final state, as shown in Fig. \ref{fig:alpha_advantage}.
If the final state is required to be uncorrelated, we observe no advantage from extra coherence in the initial state, so let us focus on the correlated final state scenario.
In the presence of coherence (\(|\alpha| \neq 0 \)) the optimal yield shows a relevant boost in a small region of \(\Delta\), while for most values of \(\Delta\) the advantage is negligible, hinting that our previous findings would not be highly impacted by having this extra resource. Finally, we note that recently other authors have studied the role of quantum correlations in photoisomerization with similar resource theoretical tools \cite{burkhard2023boosting}, however, their definition of the process yield differs from ours, making a direct comparison difficult. Despite this, our numerical procedure can also be efficiently applied to other definitions of target functions based on permutationally invariant observables.\\

{\em Discussion -- } In this work, we have quantitatively explored the impact of correlations on the efficiency of photoisomerization, using tools from quantum information theory. In particular, our results suggest that correlations between molecular switches can significantly increase the efficiency of the process. Moreover, we have studied how this boost changes when the number of subsystems becomes large, confirming that the correlation advantage vanishes in the thermodynamic limit.
To obtain these results, we have employed techniques from the resource theory of athermality, developing numerical methods and approximations that can be efficiently applied to large systems to assess state convertibility. 
As discussed in the main body, the behavior of the relative advantage provided by correlations mirrors that of a thermodynamic quantity analyzed by \cite{sapienza2019correlations}. It is still unclear how to relate the photoisomerization yield, and thus the corresponding correlation advantage, to strictly thermodynamic quantities such as the work of formation, which is surely an interesting direction for future work. \gs{Another intriguing line of inquiry would involve deploying the tools developed here to initial states that are fixed by a microscopic description of the photoisomer ensemble and its biomolecular dynamics when interacting with incoming radiation. In other words, it would be interesting to develop a microscopic model of photoexcitation and be able to infer a class of photoexcited molecular states which might exhibit collective excitations. This would provide a quantitative assessment of the impact of correlations in the initial state.}\\

{\em Acknowledgements -- }  This work was supported by the QuantERA project ExTRaQT (grant no. 499241080), the EU project C-QuENS (grant no. 101135359), and the ERC Synergy grant HyperQ (grant no. 856432).\\

{\em Data availability -- } The data that support the findings of this article are openly available \cite{data_rep}.

\bibliographystyle{unsrt}
\bibliography{ms.bib}

%\clearpage

\onecolumngrid

\newpage
\appendix

\section{Using thermomajorization curves for yield upper bounds}
\label{App:thermomajorization_curves}
\begin{figure}
    \subfloat[Thermomajorization curves of the first case $\Delta \leq \Delta_\mathrm{crit}$.\label{subfig-1:dummy}]{%
      \includegraphics[width = 0.49\textwidth]{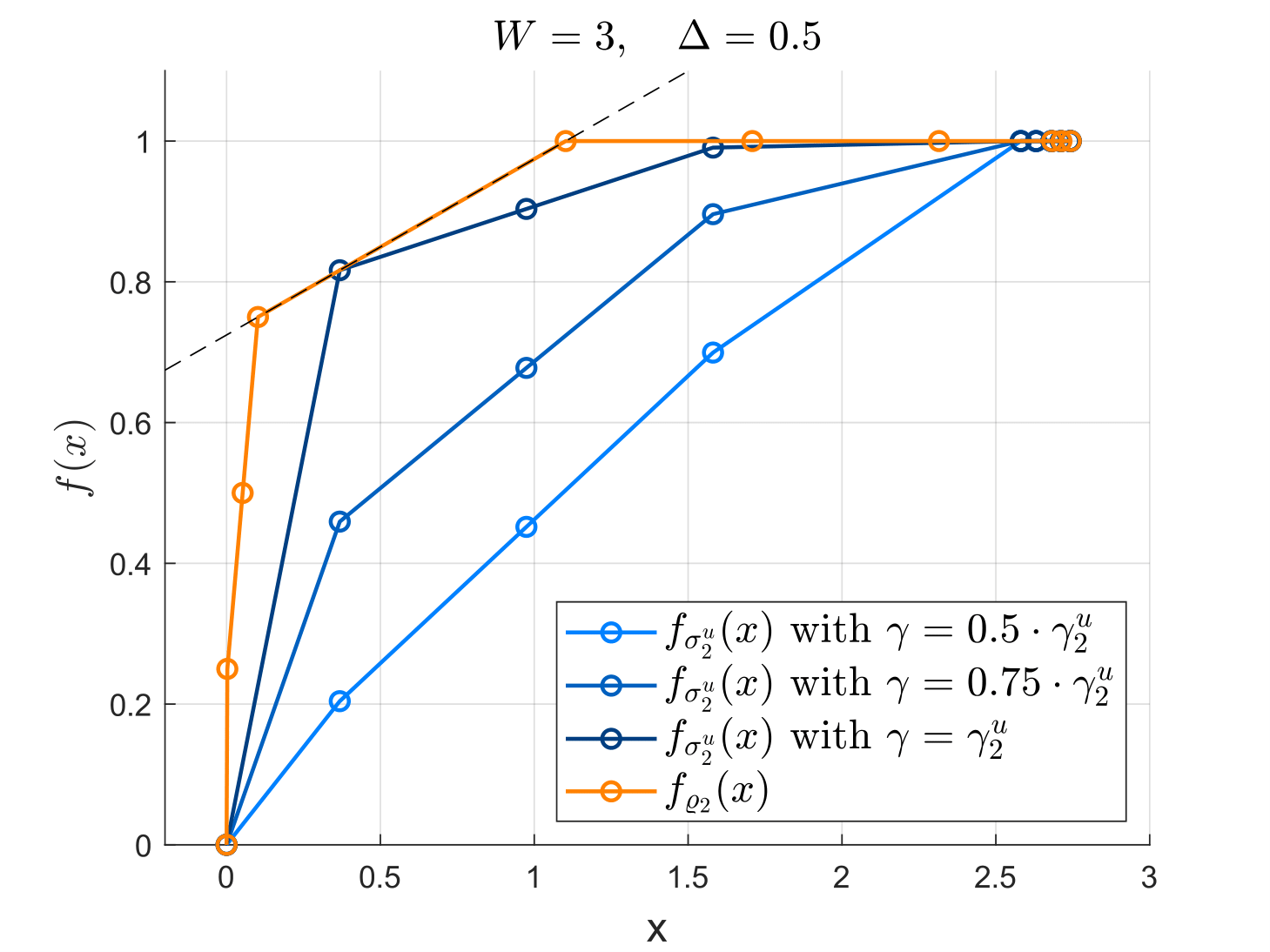}
    }
    \hfill
    \subfloat[Thermomajorization curves of the second case $\Delta \geq \Delta_\mathrm{crit}$.\label{subfig-2:dummy}]{%
      \includegraphics[width = 0.49\textwidth]{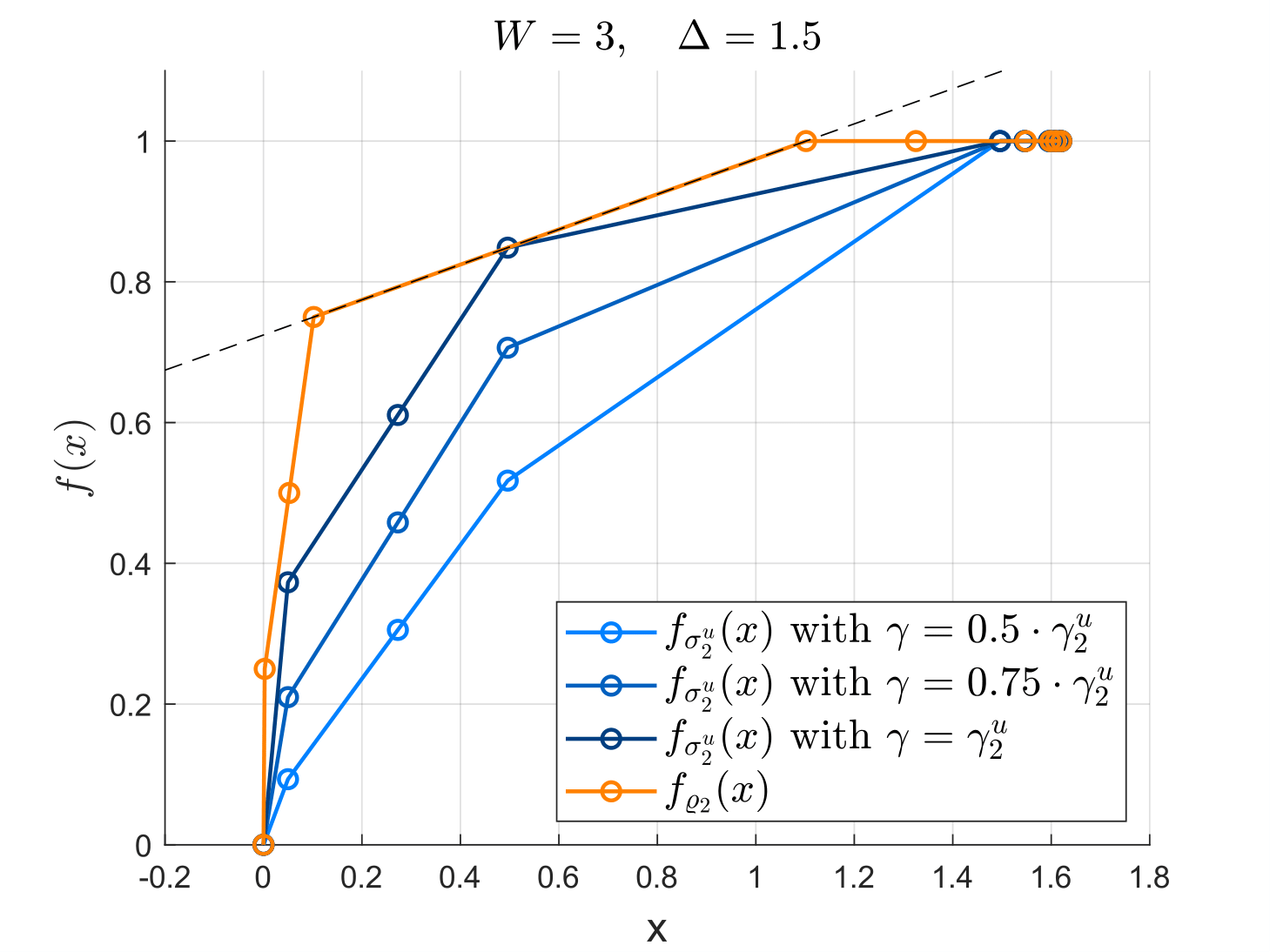}
    }
    \\
    \subfloat[Thermomajorization curves where both cases deliver the same result at $\Delta = \Delta_\mathrm{crit}$. \label{subfig-3:dummy}]{%
      \includegraphics[width = 0.49\textwidth]{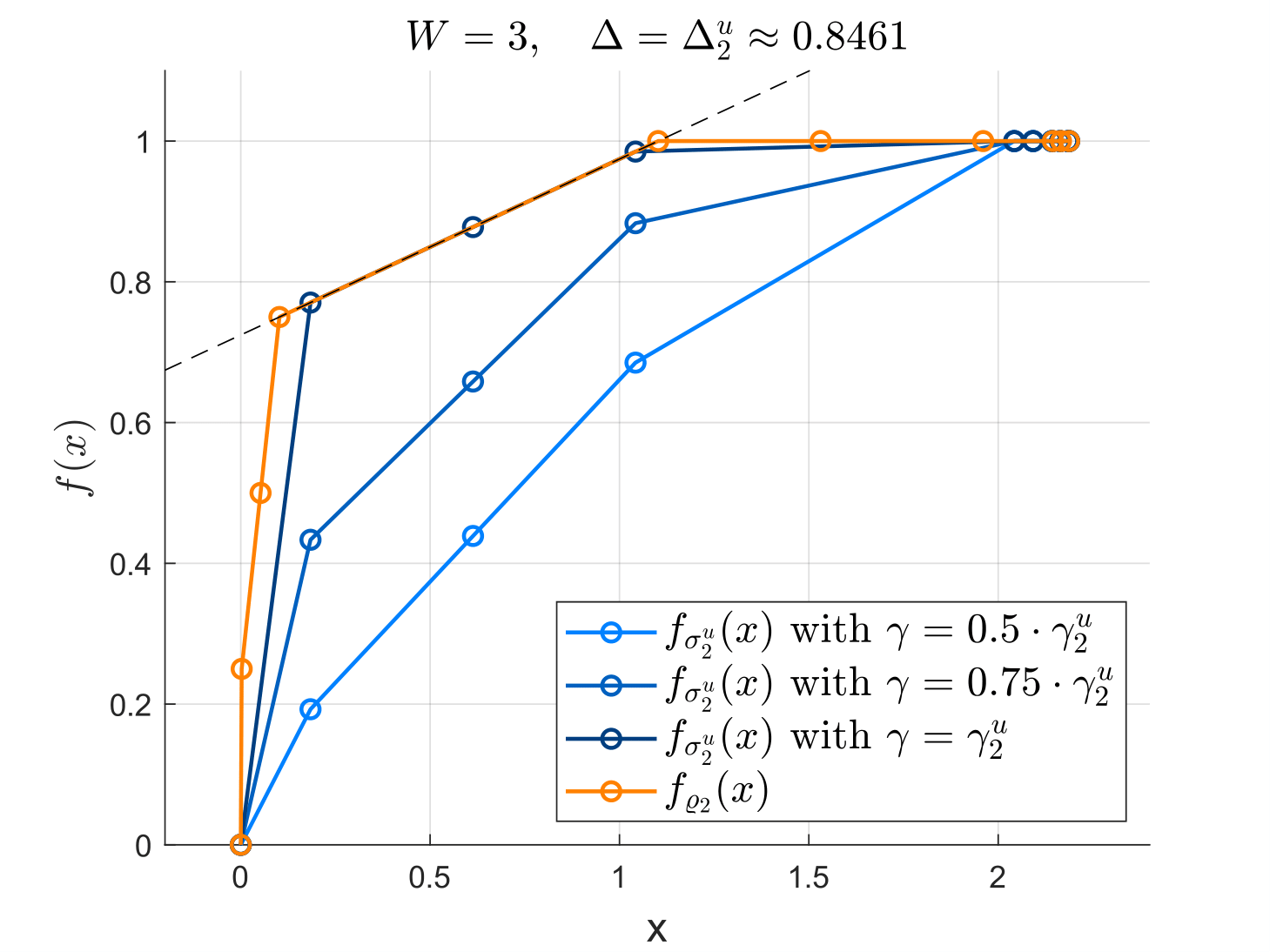}
    }
    \caption{Thermomajorization curves for different values of $\Delta$ in the setup of two molecules and an uncorrelated final state. The plots show how the final curve depends on the yield, and thus, how to maximize it under the constraints of thermomajorization. Depending on the value of $\Delta$, the final curve changes its structure and thus a case distinction is needed.}
    \label{fig:curves_two_molecules_uncorrelated}
  \end{figure}

\noindent
In this appendix, we show how the analytical form of the optimal yield is obtained from constructing thermomajorization curves. This procedure will be applied on a small number of molecules, as it becomes increasingly challenging as \(N\) increases. The construction of thermomajorization curves requires the $\beta$-ordering of the population vector \cite{lostaglio2019introductory}, which in this case is completely determined by whether the initial excitation \(q\) is smaller or larger than a threshold \(\Tilde{q} = 1/(1+e^{W})\). This defines two excitation regimes, and we focus on the high excitation case (\(q \geq \Tilde{q}\)), considering that for typical photoisomers \(W\) is large and thus \(\Tilde{q} \approx 0\). \\
Let us start with an uncorrelated final state, where we only need to optimize over a single parameter, which is the yield $\gamma$ itself. For $N=2$ molecules, one can find a depiction of the procedure used to obtain the optimal yield in Fig. \ref{fig:curves_two_molecules_uncorrelated}, which can in principle be extended to a generic number of molecules. \\
The correlated optimal yield can be calculated with the same method, with the only difference that here we will find two critical values \gs{(i.e., the values of $\Delta$ for which the $\beta$-ordering of the population vector changes, leading to elbow points in the final yield curves)} instead of one for $N=2$. Furthermore, we now have to optimize over two parameters instead of one. Generalizing this procedure to $N$ molecules then requires additional expenses, as we have to take into consideration a number of case distinctions that scales with $\mathcal{O}(N^2)$ and a number of parameters to optimize over that scales with $\mathcal{O}(N)$. Yield upper bounds can then be constructed for any arbitrary $N$ from the optimal populations according to Eq. \eqref{eq:yield_composition}, as can be seen in Fig. \ref{fig:app:numVSan} for $N=5$. \gs{We finally note that for specific molecules in realistic scenarios, the values of $\Delta$ and $W$ are known. If they are thus fixed beforehand, the procedure of finding the optimal yield is simplified and can be carried out more efficiently, since there is no need in this case to consider all possible values of $\Delta$.}

\section{The $GS_N$ Linear Program}\label{App:GSMprogram}

Gibbs-stochastic matrices can be used as an alternative to thermomajorization curves to assess state convertibility in the resource theory of athermality. While theoretically equivalent, these two methods present very different computational challenges, such as the $\beta$-ordering which is only needed for the curves. Here we show that, using Gibbs-stochastic matrices, we can construct a linear optimization problem that is efficiently solvable by standard algorithms. As we saw in Eq. \eqref{eq:optimization}, \(\gamma^c_N\) is the solution to the following optimization:
\begin{equation}
\!\max_{G \in GS_N} \Big \{\gamma(p_{\sigma_N}) \quad | \quad p_{\sigma_N} = G p_{\rho_N}  \Big \} \, .\\
\end{equation}
The entries of \(G \in GS_N\) define the amount of population transfer from one level to another. The dimension of \(G\), and thus the number of variables, scales exponentially with the number of copies. Nevertheless, we can greatly reduce the number of variables to optimize with some considerations on the states involved and the target function \(\gamma\). The following steps can be applied to obtain a reduced optimization problem where the number of variables and constraints scales only as \(\mathcal{O}(N^2)\):

\begin{enumerate}

   \item \textit{The yield depends on a subset of the parameters}: only a subset of the final state populations contribute to the yield, and only a subset of the initial state levels are populated.

   \item \textit{Populations of degenerate levels have equal weights}, resulting from the independence of \(\gamma\) from permutations of identical subsystems. We can therefore assume, without loss of generality, that any two entries of \(G\) moving population from degenerate levels to degenerate levels will be equal to each other.
\end{enumerate}

Now the target function depends only on \(N(N+1)\) distinct variables \(\{x_{i j}\}\), that is the set of entries of \(G\) moving population from levels with energy \(E = j W, j \in \{0,\dots,N\}\) to levels with energy \(E = i \Delta, i \in \{1,\dots,N\}\).\\ The constraints of the optimization still depend on entries outside of this set, but the action of entries not entering the yield can be made trivial:

\begin{enumerate}
\setcounter{enumi}{2}
    \item \textit{The Gibbs-stochasticity condition fixes an }arbitrary\textit{ column}: \(G  p_\tau = p_\tau \) fixes the parameters in one of the matrix columns \footnote{One should order the energy levels to write down the population vector of the Gibbs state, but that amounts to a permutation which is also applied on the population vector so that the expression \gs{for} the yield remains the same. An arbitrary order may be assumed.}. We find that fixing a column corresponding to an empty level in the initial state simplifies the expression of the yield, as the expression resulting from said constraints is multiplied by zero. From now on, we will fix the column of the GSM referring to energy \(E = 1\Delta\).\\ Let us call \( z_i\) the \(N\) entries of \(G\) in this column which depend on \(\{x_{i j}\}\).To be valid entries, \(z_i \in [0,1]\) must hold. This is not automatically satisfied for all choices of \(\{x_{i j}\}\) and we thus have to include them in our optimization.
    
    \item \textit{Parameters that lower the yield can be set to zero}: let us focus on the \(N\) rows of \(G\) moving population to the levels that contribute to the yield. Some of the parameters in these rows have been renamed \(x_{ij}\) and \(z_i\). We refer to the remaining ones as \(g_{ik}\), which would move population away from energy levels that are empty in the initial state. We call \(\mathcal{S}\) the set of column indices selecting these entries of \(G\). With this notation, the expression of the parameters $z_i$ reads
    
    \begin{equation}
    z_i = \frac{e^\Delta}{N} \left [ e^{-i \Delta} - \sum_{j=1}^{N+1}\binom{N}{j-1}x_{i j} e^{-(j-1) W} - \sum_{k \in \mathcal{S}} \text{deg}(k) g_{i k} e^{-E_k}  \right ] \, ,
    \end{equation}
    
    where we indicated with deg($k$) the number of degenerate levels with energy \(E_k\). It is trivial to show that \(z_i \leq 1 \) is always true, as the entries of \(G\) are all positive. The condition \(0 \leq z_i \) reads
    
    \begin{equation}
    \sum_{j=1}^{N+1}\binom{N}{j-1}x_{i j} e^{-(j-1) W} \leq  e^{-i \Delta}  - \sum_{k \in \mathcal{S}} \text{deg}(k) g_{i k} e^{-E_k}.
    \end{equation}
    
    This inequality poses a constraint on a linear combination of $x_{i j}$ with positive coefficients. Since the yield depends on \(\{x_{ij}\}\) via positive coefficients, we want this upper bound to be as high as possible. Thus, we set to zero all the coefficients $g_{ik}$, obtaining:
    
    \begin{equation}
    z_i = \frac{e^\Delta}{N} \left [ e^{-i \Delta} - \sum_{j=1}^{N+1}\binom{N}{j-1}x_{i j} e^{-(j-1) W} \right ].
    \end{equation}
\end{enumerate}

The resulting optimization problem reads

\noindent\begin{minipage}{\linewidth}
    \begin{maxi}|l|
	  {x_{i j}}{\gamma = \sum_{i=1}^N \binom{N}{i} \frac{i}{N} p_i}{}{}
        \addConstraint{x_{i j} \in [0,1], \quad}{i=1 \ldots N, \quad j=1 \ldots N+1}
        \addConstraint{z_{i} \in [0,1], \quad}{i=1 \ldots N}
   	\addConstraint{\sum_{i=1}^N \binom{N}{i} x_{i j} \leq 1, \quad}{j=1 \ldots N+1}
   	\addConstraint{\sum_{i=1}^N \binom{N}{i} z_i \leq 1, \quad}{\quad}
     \end{maxi}

\begin{equation*}
p_i = \sum_{j=1}^{N+1}\binom{N}{j-1}x_{i j} (1-q)^{N-(j-1)}q^{j-1},
\end{equation*}

\begin{equation*}
z_i = \frac{e^\Delta}{N} \left [ e^{-i \Delta} - \sum_{j=1}^{N+1}\binom{N}{j-1}x_{i j} e^{-(j-1) W}  \right ].
\end{equation*}
\end{minipage}
\vspace{1cm}

\begin{figure}[ht]
    \centering
    \begin{minipage}[b]{0.48\textwidth}
        \centering
        \includegraphics[width=\linewidth]{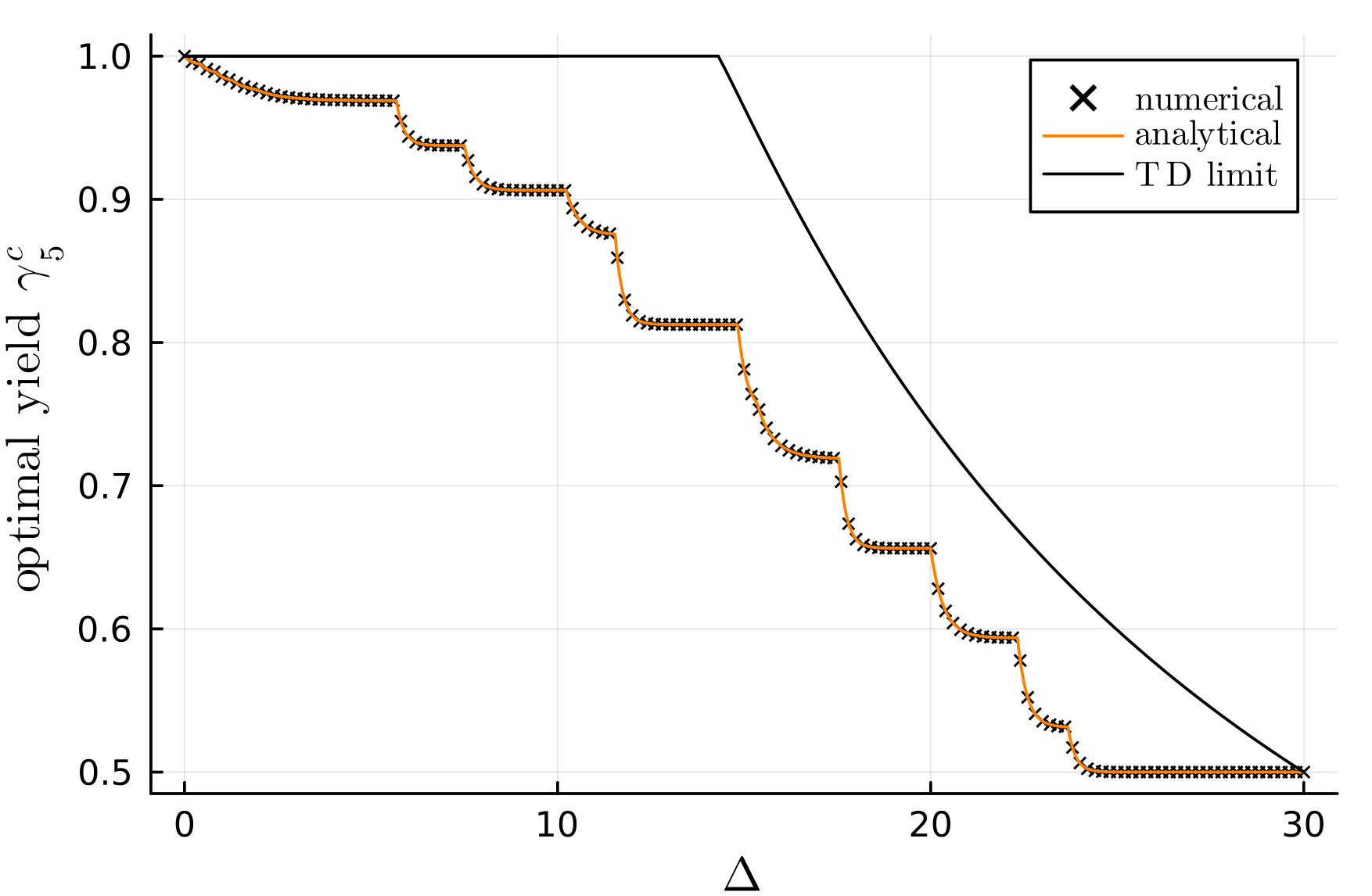}  % Adjust path
        \caption{Numerical optimization (cross markers) compared with the analytical solution (solid line) for \(N=5\) molecules. The match between the two \gs{supports} the reliability of the numerical approach.}
        \label{fig:app:numVSan}
    \end{minipage}
    \hfill
    \begin{minipage}[b]{0.48\textwidth}
        \centering
        \includegraphics[width=\linewidth]{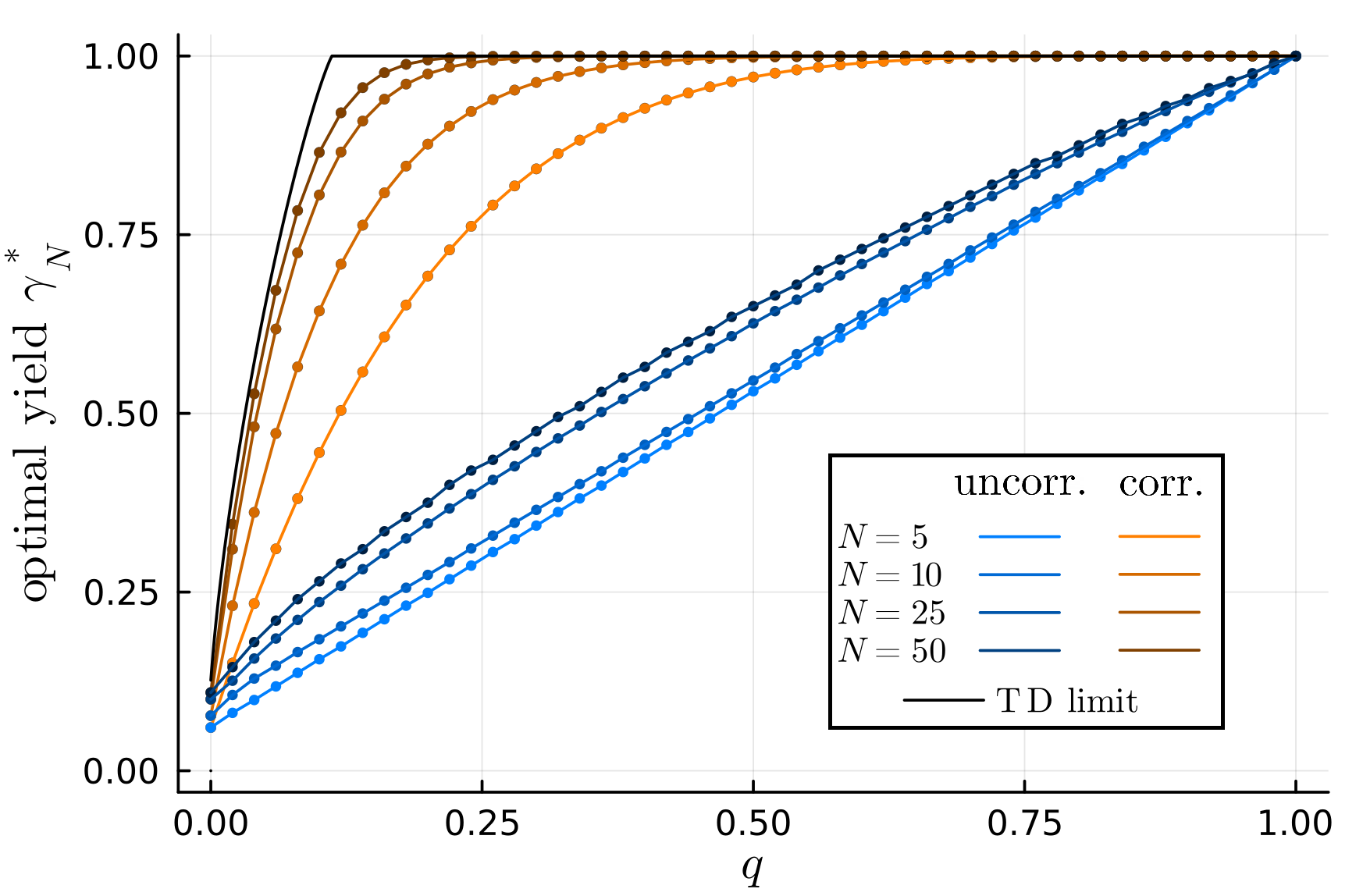}  % Adjust path
        \caption{Optimal yield with uncorrelated (blue) and correlated (orange) final states, as a function of the initial state population \(q\) for a molecule with energies \(\Delta = 3\) and \(W = 30\).}
        \label{fig:app:Gammaq}
    \end{minipage}
\end{figure}

A comparison between the numerical optimization results and the analytical solution for a small system, for which the thermomajorization curves method is still feasible, is shown in Figure \ref{fig:app:numVSan}. In this paper, we conducted our analysis as a function of \(\Delta\), in continuity with literature. However, this optimization can also be performed as a function of other system parameters, such as \(q\), with no change needed. Figure \ref{fig:app:Gammaq} shows similar findings to figure \ref{fig:GammaDelta} while varying \(q\) and keeping \(\Delta\) fixed.

\section{The $GS_N$ Linear Program in the large $N$ \gs{regime}}
\label{App:GSMasymptotic}

The populations of the initial state \(\rho^{\otimes N}\) are distributed on sets of degenerate levels as described in Eq. \eqref{eq:popdistr}. When \(N \rightarrow \infty\) the state can be approximated by its most populated levels. In particular, the population distribution in this \sout{limit} \gs{regime} peaks on energy levels with energy \(q N W\):
\begin{equation} \rho ^ {\otimes N} \approx \binom{N}{qN}^{-1} \sum_{ij} P_{i j} \ket{0^{\otimes (1-q)N} W^{\otimes qN}} \bra{0^{\otimes (1-q)N} W^{ \otimes qN}} P_{i j}^{\dagger}  \end{equation}
%Using the Stirling approximation we find that 
%\begin{equation} 
%\binom{N}{qN} \approx \exp \left \{N \left [ H(\rho_{i}) - %\frac{1}{2N}\ln{[2 \pi N q (1-q)]} \right] \right \}
%\end{equation}
%the correction on the entropy vanishing as $N \rightarrow \infty $. 
where \(P_{ij}\) permutes a pair of molecules. With this approximation, the entire population of the initial state is distributed on one set of degenerate energy levels. Repeating the same reasoning of Appendix \ref{App:GSMprogram}, there will be only $N$ distinct transfer rates $x_i$, one for each target level on the final state with energy \(i \Delta\) with \(i \in \{ 1,\dots, N \} \).  The simplified optimization problem now reads as follows:
\noindent\begin{minipage}{\linewidth}
    \begin{maxi}|l|
	  {x_i}{\gamma = \sum_{i=1}^N \binom{N}{i} \frac{i}{N} x_i}{}{}
    \addConstraint{x_i}{\in [0,1], \quad}{i=1\ldots N}
   \addConstraint{\sum_{i=1}^N \binom{N}{i} x_i}{\leq 1 }{}
	  \addConstraint{x_i}{\leq \exp{- i \Delta + N \Delta^*}\quad}{i=1\ldots N} \, .
   \label{asympt_opt}
     \end{maxi}
\end{minipage}

\vspace{1cm}
Interestingly, the free energy of the initial state \(\Delta^* = qW +q\ln{q} + (1-q)\ln{(1-q)} \) appears now explicitly. If we expand the sum defining $\gamma$, the parameter $x_N$ always appears with a multiplying coefficient equal to 1:
\begin{equation} \gamma = x_1 + (N-1) x_2 + \ldots + (N-1) x_{N-1} + x_N \, , \end{equation}
while the constraints acting on this transfer rate read 
\begin{align}
x_N &\leq 1 - \sum_{i=1}^{N-1} \binom{N}{i} x_i \, ,\\
x_N &\leq \exp{-N (\Delta - \Delta^*)} \, .   
\end{align}
If \(\Delta \leq \Delta^*\) the right-hand side of the second constraint is always greater than \(1\), the constraint is therefore not active. This defines a range of $\Delta$ in which we can set $x_N = 1$ and all other $x_i = 0$, thus obtaining $\gamma^c_\infty(\Delta \leq \Delta^*) \equiv 1$.

\section{Refining the model}\label{sec:refining}

\begin{figure}[ht]
    \centering
    \includegraphics[width=0.3\linewidth]{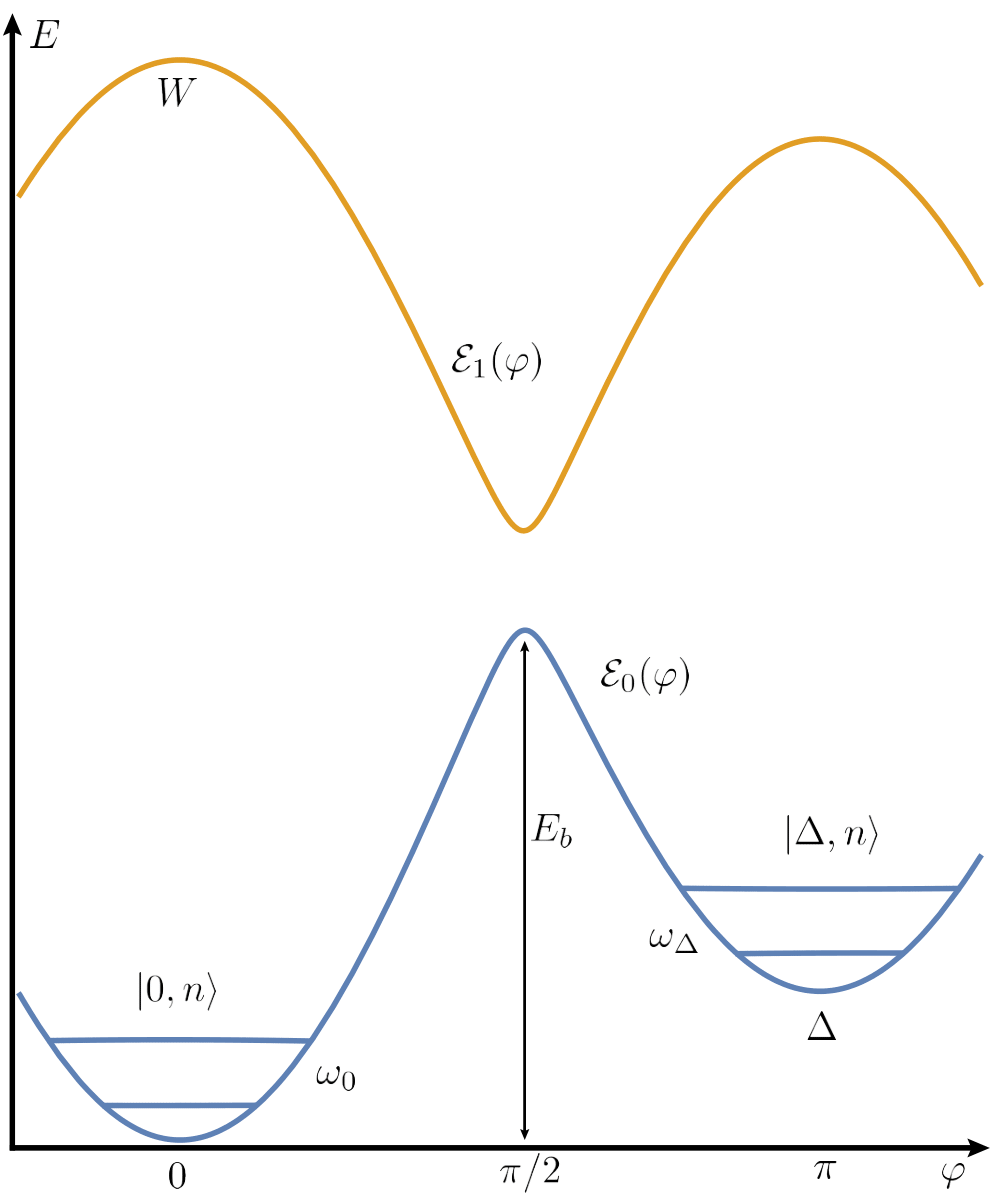}
    \hspace{1cm}
    \includegraphics[width=0.5\linewidth]{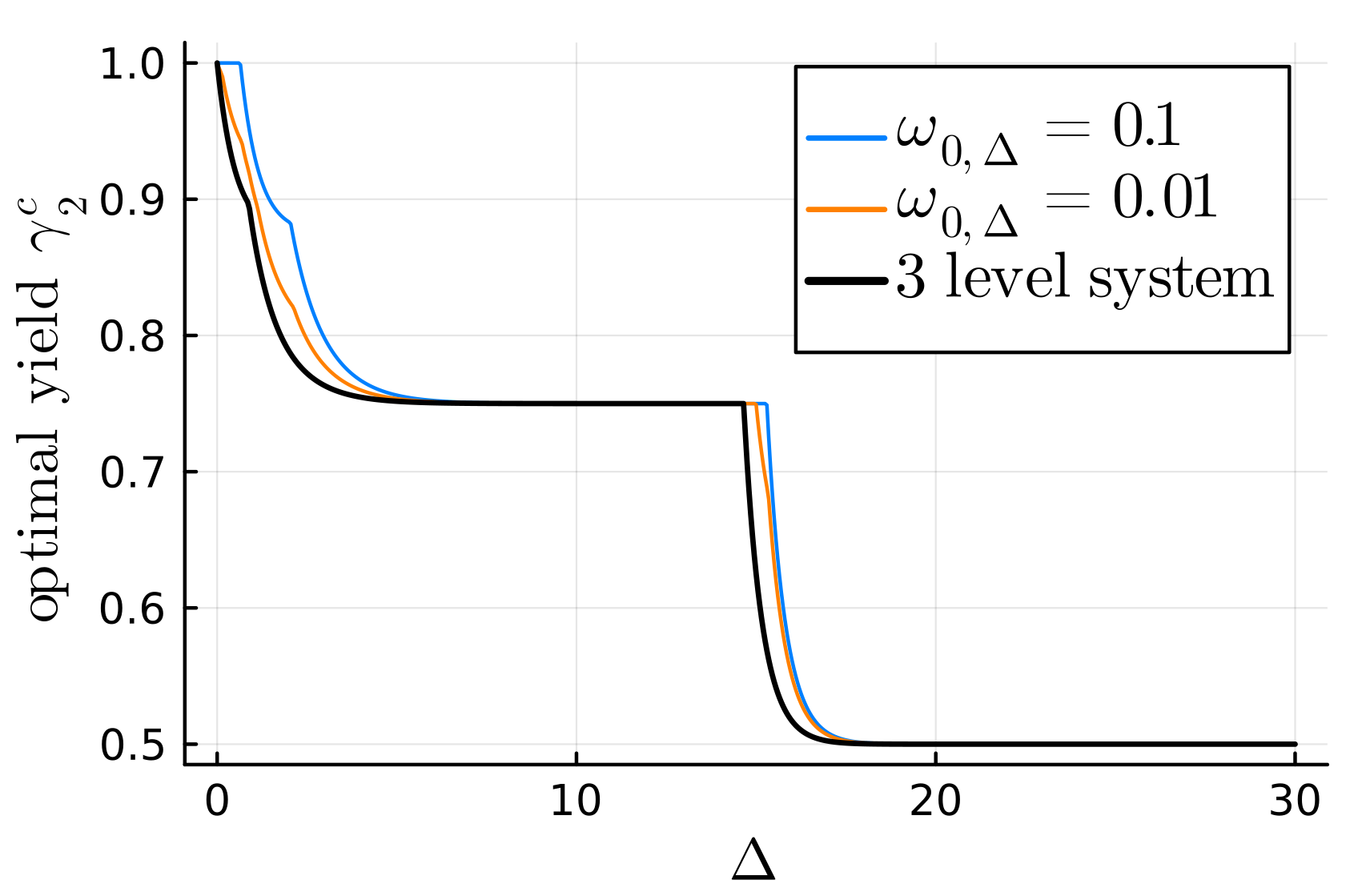}
    \caption{Left: 5-level system model, obtained by considering one extra level for each minimum of the electronic ground state \gs{(the spacing between levels is exaggerated for legibility and is not in scale)}. Right: Optimal efficiency for \(N=2\) photoswitches, each modeled with a 5-level system\gs{, with \(\Delta\in [0,W=30]\)}. Correlations are allowed in the final state, and the optimal efficiency is calculated for two different values of \(\omega_{0,\Delta}\). The bounds are obtained for a high local inverse temperature \(\beta_0 = 100\), consistent with the Hilbert space truncation. The three-level model is shown for comparison.}

    \label{fig:5L}

\end{figure}

Following the previous literature on the description of photoisomerization through the resource theory of athermality, the Hilbert space of the electronic degrees of freedom of the rotating molecule was truncated to a 3-level system. This will impact the thermodynamic behavior of the system, which has fewer degrees of freedom available to spread and raise its entropy. The dimensionality of the Hilbert space can indeed be considered a thermodynamical resource (see i.e. \cite{silva2016performance}), but it is not immediately clear what will the magnitude of this advantage be. In theory, one should account for the full vibrational spectrum of the molecule, but this would quickly become numerically intractable. Instead, we keep the assumption of fairly localized states in \(\varphi\) and improve the model's resolution by introducing one extra level in each potential well in the energy landscape. Following the arguments in \cite{tiwary2024quantum}, this results in a 5-level model, as shown in Fig. (\ref{fig:5L}). We can now use the numerical method based on the $GS_N$ set developed in this paper to explore the effect of these extra levels. The order of magnitude of the energy of these new first excited states \(\omega_0\) and \(\omega_\Delta\) are, in real molecules, much smaller than $\Delta$ and $W$ (see \cite{chuang2022steady}). To simplify the calculations, we choose them to be equal and obtain the efficiency bounds for \(\omega_{0,\Delta} \in \{0.1,0.01\}\).
%of magnitude \(10^{-1} - 10^{-2} K_bT\) 

The initial state \(\rho\) must now be redefined, as the fraction of population \((1-q)\) which was not excited by the incident radiation must now be distributed on two energy levels. On these two levels, we impose a local thermalization, so that the total population \((1-q)\) is distributed as:
\begin{align}
    q_0 &= (1-q) \frac{1}{1+\exp(-\beta_{0} \omega_0)} \, ,\\
    q_{\omega_0} &= (1-q) \frac{\exp(-\beta_{0} \omega_0)}{1+\exp(-\beta_{0} \omega_0)} \, ,
\end{align}
with \(\beta_0\) being the effective inverse temperature on the subsystem. To maintain our Hilbert space truncation approximation consistent, we choose a large \(\beta_0\), so that the population distribution peaks at the lowest level. 
In redefining Eq. \eqref{observable}, we impose that levels at \(\varphi = \pi\) contribute equally regardless of their energy :
\begin{equation}
    \mathcal{O}  = \frac{1}{N} \sum_{n}\sum_{i=1}^N (|\Delta,n \rangle \langle \Delta,n | )_i \, . 
\end{equation}
As we can see from Fig. \ref{fig:5L}, the presence of extra levels can indeed increase the maximum yield, by allowing populations to spread on more levels contributing to the yield (at least, this is the case for some regions of \(\Delta\)).
%dump:
%\input{D0}
%\input{Multiple_ps}
%\input{large_N}

\newpage
\twocolumngrid
\end{document}